\documentclass[useAMS,usenatbib]{mn2e}

\usepackage{graphics}
\usepackage{graphicx}
\usepackage{amsmath}
\usepackage{epsf}
\usepackage{gensymb}
\usepackage{rotating}
\usepackage{array}
\usepackage{subfigure}
\usepackage{longtable}
\usepackage{hyperref}
\usepackage{epstopdf}
\usepackage{lscape}
\usepackage{url}

\newcolumntype{C}[1]{ > {\centering\let\newline\\\arraybackslash\hspace{0pt}}m{#1}}
\newcolumntype{L}[1]{ > {\let\newline\\\arraybackslash\hspace{0pt}}m{#1}}

\def\reff@jnl#1{{\rm#1\/}}

\def\aj{\reff@jnl{AJ}} % Astronomical Journal
\def\araa{\reff@jnl{ARA\&A}} % Annual Review of Astron and Astrophys
\def\apj{\reff@jnl{ApJ}} % Astrophysical Journal
\def\apjl{\reff@jnl{ApJ}} % Astrophysical Journal, Letters
\def\apjs{\reff@jnl{ApJS}} % Astrophysical Journal, Supplement
\def\ao{\reff@jnl{Appl.Optics}} % Applied Optics
\def\apss{\reff@jnl{Ap\&SS}} % Astrophysics and Space Science
\def\aap{\reff@jnl{A\&A}} % Astronomy and Astrophysics
\def\aapr{\reff@jnl{A\&A~Rev.}} % Astronomy and Astrophysics Reviews
\def\aaps{\reff@jnl{A\&AS}} % Astronomy and Astrophysics, Supplement
\def\azh{\reff@jnl{AZh}} % Astronomicheskii Zhurnal
\def\baas{\reff@jnl{BAAS}} % Bulletin of the AAS
\def\jrasc{\reff@jnl{JRASC}} % Journal of the RAS of Canada
\def\memras{\reff@jnl{MmRAS}} % Memoirs of the RAS
\def\mnras{\reff@jnl{MNRAS}} % Monthly Notices of the RAS
\def\pra{\reff@jnl{Phys.Rev.A}} % Physical Review A: General Physics
\def\prb{\reff@jnl{Phys.Rev.B}} % Physical Review B: Solid State
\def\prc{\reff@jnl{Phys.Rev.C}} % Physical Review C
\def\prd{\reff@jnl{Phys.Rev.D}} % Physical Review D
\def\prl{\reff@jnl{Phys.Rev.Lett}} % Physical Review Letters
\def\pasp{\reff@jnl{PASP}} % Publications of the ASP
\def\pasj{\reff@jnl{PASJ}} % Publications of the ASJ
\def\qjras{\reff@jnl{QJRAS}} % Quarterly Journal of the RAS
\def\skytel{\reff@jnl{S\&T}} % Sky and Telescope
\def\solphys{\reff@jnl{Solar~Phys.}} % Solar Physics
\def\sovast{\reff@jnl{Soviet~Ast.}} % Soviet Astronomy
\def\ssr{\reff@jnl{Space~Sci.Rev.}} % Space Science Reviews
\def\zap{\reff@jnl{ZAp}} % Zeitschrift fuer Astrophysik
\def\nat{\reff@jnl{Nature}} % Nature
\def\icarus{\reff@jnl{Icarus}} % Icarus
\def\acta{\reff@jnl{Acta Astron.}} % Acta Astronomy

\hypersetup{colorlinks=true,linkcolor=blue,citecolor=blue,filecolor=blue,urlcolor=blue}

\title[Besan\c{c}on model analysis of MOA-II microlensing]{Besan\c{c}on Galactic model analysis of MOA-II microlensing: evidence for a mass deficit in the inner bulge\thanks{This work uses the Manchester--Besan\c{c}on Microlensing Simulator - MaB$\mu$lS, which is publicly available online at \url{http://www.mabuls.net/}}}
\author[S. Awiphan, E. Kerins and A. C. Robin]{S. Awiphan$^{1}$\thanks{E-mail: supachai.awiphan@student.manchester.ac.uk}, E. Kerins$^{1}$ and A. C. Robin$^{2}$\\
$^{1}$Jodrell Bank Centre for Astrophysics, School of Physics and Astronomy, University of Manchester, Oxford Road, Manchester M13 9PL, UK\\
$^{2}$Institut Utinam, CNRS UMR6213, Universit{\'e} de Franche-Comt{\'e}, OSU THETA Franche-Comt{\'e}-Bourgogne, BP 1615, Besan\c{c}on 25010, France}
\begin{document}

\date{}

\pagerange{\pageref{firstpage}--\pageref{lastpage}} \pubyear{2002}

\maketitle

\label{firstpage}

\begin{abstract}
Galactic bulge microlensing surveys provide a probe of Galactic structure. We present the first field-by-field comparison between microlensing observations and the Besan\c{c}on population synthesis Galactic model. Using an updated version of the model we provide maps of optical depth, average event duration and event rate for resolved source populations and for difference imaging (DIA) events. We also compare the predicted event timescale distribution to that observed. The simulation follows the selection criteria of the MOA-II survey \citep{sum2013}. We modify the Besan\c{c}on model to include M dwarfs and brown dwarfs. Our best fit model requires a brown dwarf mass function slope of $-0.4$. The model provides good agreement with the observed average duration, and respectable consistency with the shape of the timescale distribution (reduced $\chi^2 \simeq 2.2$). The DIA and resolved source limiting yields bracket the observed number of events by MOA-II ($2.17\times$ and $0.83\times$ the number observed, respectively). We perform a 2-dimensional fit to the event spatial distribution to predict the optical depth and event rate across the Galactic bulge. The most serious difficulty for the model is that it provides only $\sim 50\%$ of the measured optical depth and event rate per star at low Galactic latitude around the inner bulge ($|b|<3{^\circ}$). This discrepancy most likely is associated with known under-estimated extinction and star counts in the innermost regions and therefore provides additional support for a missing inner stellar population. 
\end{abstract}


\section{Introduction}

The microlensing surveys toward the Galactic bulge have provided useful information for the search for exoplanets and for the study of Galactic structure \citep{pac1996,gau2012}. Several microlensing surveys have monitored a large number of stars and detected thousands of events over the bulge [e.g. OGLE \citep{uda1994,sum2006,wyr2015}, MOA \citep{sum2003,sum2013}, MACHO \citep{alc1997,alc2000,pop2005} and EROS \citep{afo2003,ham2006}]. The microlensing optical depth, $\tau$, measures the fraction of the sky covered by the Einstein rings of the lenses for a given line of sight. As the optical depth is directly related to the mass density of the lens population, it can be used to determine the mass distribution of the bulge. However, a difficulty in measuring the microlensing optical depth stems from the fact that it is sensitive to the individual contributions of long duration events. Another measurable property from the surveys is microlensing event rate, $\Gamma$, which has the advantage that it is not dominated by a small number of long duration events but the disadvantage that it is sensitive to Galactic kinematics and the stellar mass function, as well as the mass distribution.

A number of measurements of the bulge optical depth have been made by the survey teams, often under different sample definitions. We loosely categorize these as: resolved source measurements, difference image analysis (DIA) source measurements and red clump giant (RCG) source measurements. The resolved source method includes all sources which are brighter than magnitude limit, whilst the DIA method includes fainter sources which may only be detectable during lensing. The DIA method has the benefit that it is less sensitive to blending systematics within crowded fields and potentially provides a better S/N ratio measurement due to the larger available sample size. At the other extreme the RCG method uses samples of events which involve only bright sources which are assumed to be well resolved and therefore should exhibit a minimal blending bias. In recent studies, DIA optical depth measurements tend to be about 25\% higher that those derived from RCG samples \citep{sum2013}.

The MOA-II survey \citep{sum2013} determined the optical depth from a study of 474 events with sources brighter than 20$^{th}$ magnitude in the {\it I}-band toward the bulge. They determined a value of $\tau_{\textup{DIA}} = [2.35\pm 0.18]e^{[0.51\pm 0.07](3-\left | b \right |)}\times 10^{-6}$~$^[$\footnote{The subscripts of optical depths indicate the method of analysis and the long duration cutoff in days.}$^]$. For the average optical depth, they find that $\tau_{\textup{DIA},200} = 3.64_{-0.45}^{+0.51}\times 10^{-6}$ at ($l = 0.97^{\circ}, b = -2.26^{\circ}$). These results are broadly consistent with previous measurements from MOA-I \citep{sum2003}, OGLE \citep{sum2006}, MACHO \citep{pop2005} and EROS \cite{ham2006} (See Table~\ref{OMlist}).

Over recent years, more detailed theoretical models have been developed in order to predict the microlensing optical depth values \citep{han2003,woo2005,ker2009}. \cite{ker2009} presented synthetic maps of optical depth and event rate over the Galactic bulge using catalogues generated from the Besan\c{c}on galactic model developed by \cite{rob2003} with 3D extinction maps from \cite{mar2006}. The observational result tends to agree with the theoretical models. However, the recent MOA-II surveys provide optical depth of RCG 30-40\% below the prediction of \cite{ker2009} which might be the result of lacking long crossing time events in observational data \citep{sum2013}. 

One important issue which we do not explicitly address is source blending. In principle this can be examined within the context of a population synthesis model through construction of artificial images. However, this is beyond the scope of the present work. Instead, we choose to model only the two idealized cases described above (resolved and DIA sources). If the model is a good representation of reality, these cases should provide reasonable upper and lower limits on the potential number of events within specific MOA-II sources sub-sample. Our DIA estimate should always provide a firm upper limit to the observed microlensing rate per star. On the other hand, our resolved source calculations should provide a firm lower limit to the number of events per star. MOA-defined sub-samples such as clump giants should yield a rate somewhere intermediate to these regimes as the RCG sources are resolved but known to be confined to the bulge, whereas in our simulation a non-negligible number of our sources will be closer to the observer. We note that, in order for the model to accurately compute the microlensing rate per unit sky area it would also be necessary to accurately mimic the source colour-magnitude cuts of the survey.

In this paper, the microlensing optical depth and event rate maps are presented by using a recent version of Besan\c{c}on galactic model \citep{rob2014} and compared to the MOA-II result. The updated model includes an inner bar component. In Section 2, the Besan\c{c}on Galactic model is summarised. In order to simulate the MOA-II microlensing event sample from the Besan\c{c}on model, the selection criteria is discussed in Section 3. The calculation method of microlensing parameters and their maps are shown in Section 4. In Section 5, the results from the Besan\c{c}on simulation are compared with the observational results of the MOA-II survey. The model parametrisation of simulation results are provided in Section 6. Finally, in Section 7, the conclusions of this work are presented.

\section{The Besan\c{c}on Galactic model}

The Besan\c{c}on model, a Galactic population synthesis model, is designed to describe the observable properties of the Galactic stellar population by relating them to models of Galactic formation and evolution, stellar formation and evolution and stellar atmospheres, using constraints from observation data \citep{rob2003,rob2012,rob2014}. In the Besan\c{c}on model, stars are created from gas following an initial mass function (IMF) and star formation rate (SFR), and evolved according to theoretical stellar evolutionary tracks. For each simulated star, the photometry, kinematics and metallicity are computed. In order to simulate the Galaxy, four main populations are assumed: a thin disc; thick disc; bulge/bar; and stellar halo.

The model also includes a 3D extinction map \citep{mar2006}. An interstellar extinction distribution in three dimensions from 2MASS survey \citep{cut2003} towards the inner Galaxy ($|l| \leq 100^{\circ}$ and $|b| \leq 10^{\circ}$), with 15$^\prime$ resolution is used. \cite{mar2006} calculated the extinction as a function of distance along each line of sight by comparing observed reddened stars to unreddened simulated stars from the Besan\c{c}on model. This distribution can be used to determine the observed colours and magnitudes of the simulated stars. In the following work, a later version of Besan\c{c}on model \citep{rob2014} has been used.

\subsection{Thin disc}

The thin disc is a major component in the Galactic central region. It is assumed to have an age of 10 Gyr. A constant SFR over the past 10 Gyr is assumed, along with an IMF with two slopes, $\textup{d} N/\textup{d}m \propto M^{-1.6}$ for $M<1M_{\odot}$ and $\textup{d} N/\textup{d}m \propto M^{-3.0}$ for $M>1M_{\odot}$. The total mass of the thin disk is $9.3\times10^{9}~M_{\odot}$ . The luminosity function determined from {\it Hipparcos} observations is adopted \citep{hay1997a,hay1997b,rob2003}, whilst the underlying density law follows the \cite{ein1979} density profile. The disc is modelled with a central hole and so the maximum density of the thin disc is located at about 2.5 kpc from the Galactic Centre. The kinematics follow the {\it Hipparcos} empirical estimates of \cite{gom1997}. The populations of thin disc are divided into 7 distinct components with different distribution in age, scale height and velocities \citep{rob2012}.

\subsection{Thick disc}

The thick disc is of much lower density than the thin disc locally but becomes important at Galactic latitudes above about 8-10$^{\circ}$. In the model it is assumed a separate population from the thin disc, with distinct star formation history. Recent constraints from SDSS and 2MASS data lead to revisions of the scale length and scale heights \citep{rob2014}. We here make use of the single thick disc episode of formation presented in \cite{rob2014}, modelled by a 12 Gyr isochrone of metallicity -0.78 dex, with a density law following a modified exponential (parabola up to z = 658 pc, followed by an exponential with a scale height of 533 pc), which is roughly equivalent to a sech$^2$ function of scale height 450 pc. The radial density follows an exponential with a scale length of 2.355 kpc. Its kinematics follow the result of \cite{ojh1996}.

\subsection{Bulge/bar}

A new model of the bulge of the Besan\c{c}on model has been proposed by \cite{rob2012}, as the sum of two ellipsoids: a standard boxy bulge (bar), the most massive component which dominates the stellar content of latitudes below about 5$^{\circ}$, and another ellipsoid (thick bulge) with longer and thicker structure which can be observe at higher latitudes where the bar starts to be less prominent. However, in \cite{rob2014}, we showed that the ``thick bulge'' population was in fact the inner part of the thick disc which short scale length makes a large contribution in the bulge region. Hence, in this new version, the populations in the bulge region are: the thin disc, the bar and the thick disc. The angle of the bar to the Sun-Galactic Centre direction is 13$^{\circ}$. The bar kinematics are taken from the model of \cite{fux1999} and the bulge kinematics are established to reproduce the BRAVA survey data \citep{ric2007}. The stellar density and luminosity function are assumed from the result of \cite{pic2004} with a single burst population of 10 Gyr age. The IMF below and above 0.7~$M_{\odot}$ are assumed to be $\textup{d} N/\textup{d}m \propto m^{-1.5}$ and a Salpeter slope, $\textup{d} N/\textup{d}m \propto m^{-2.35}$, respectively \citep{pic2004}. The total bar mass is $5.9\times10^{9}~M_{\odot}$. The model mass to light ratio is 2.0 at the Sagittarius Window Eclipsing Extrasolar Planet Search (SWEEPS) field ($l=1.25^{\circ},b=-2.65^{\circ}$) in Johnson-I band which is compatible with result of \cite{cal2015} in F814W filter (wide $I$).

\subsection{Stellar halo}

The stellar halo is older than the thick disc (14 Gyr) and metal poor ($\left [\textup{Fe}/\textup{H} \right ]$ = -1.78). A single burst population with an IMF, $\textup{d} N/\textup{d}m \propto m^{-0.5}$, and total mass of $4.0\times10^{10}~M_{\odot}$ are assumed \citep{rob2003}. The density law has been revised in the study of SDSS+2MASS star counts \citep{rob2014}. It is now modelled with a power law density with an exponent of 3.39 and an axis ratio of 0.768. Its kinematics is modelled with Gaussian distributions of velocities of dispersion (131, 106, 85) in km/s in the (U,V,W) plane, and no rotation.

\section{Simulating the Galactic bulge}

\subsection{Simulating the MOA-II fields}
\label{sec:sim}

In following work, we simulate the MOA-II survey data taken from the 2006 and 2007 observing seasons \citep{sum2011,sum2013}. In order to obtain enough samples in each magnitude range, we produce lens/source star catalogues spanning four $H$-band magnitude ranges. $H$-band selection ensures that we adequately samples all relevant stellar types, though we stress that our calculations are performed using the corresponding $R$ and $I$-band magnitudes of the sources since these are the relevant filters for MOA-II. Our ranges correspond to: $-10\leq H<15$, $15\leq H<19$, $19\leq H<23$ and $H>23$. The latter ranges are dominated by stars which are too faint to act as sources but which do act as lenses. The solid angle in each catalogue, $\Omega_{\textup{sim}}$, is chosen to contain $\sim$6,000 stars in each range towards Baade's Window $(l=1^{\circ},b=-4^{\circ})$ (Table~\ref{catalogue}). The first catalogue ($-10\leq H<15$) has a solid angle of 0.026 deg$^{2}$, corresponding to the size of the MOA-II sub-fields. The simulation catalogues stars out to a distance of 15 kpc and has the same overall areal coverage as the MOA-II survey. Our final results are appropriately inverse weighted with $\Omega_{\textup{sim}}$ in order to recover the relevant microlensing observables.

\begin{table}
\caption{Solid angles used for the simulated catalogues. $\Omega_{\textup{sim}}$ is used to compute the spatial maps presented in the paper, whilst $\Omega_{\textup{sample}}$ is used to compute the global time scale distribution (See text).}
\centering
{\footnotesize 
\begin{tabular}{ccc}
\\
\hline\hline
\textbf{Magnitude range} & \textbf{$\Omega_{\textup{sim}}$ (deg$^{2}$)} & \textbf{$\Omega_{\textup{sample}}$ (deg$^{2}$)} \\ [3pt]
\hline
$ -10\leq H<15$ & $2.6\times 10^{-2}$ & $4.5\times 10^{-3}$ \\[1pt]
$ 15\leq H<19$ & $8.4\times 10^{-4}$ & $1.4\times 10^{-4}$ \\[1pt]
$ 19\leq H<23$ & $9.6\times 10^{-5}$ & $1.6\times 10^{-5}$ \\[1pt]
$ H>23$ & $1.2\times 10^{-4}$ & $2.0\times 10^{-5}$ \\[1pt]
\hline
\end{tabular}
}
\label{catalogue}
\end{table}

For each line of sight, the microlensing optical depth, average time scale and event rate toward the Galactic bulge are calculated using all combinations of source and lens pairs from the four catalogues. We compute microlensing quantities obtained from all resolved sources above a specific magnitude threshold and also from all difference imaging analysis (DIA) sources which have a magnified peak above the same threshold \citep{ala2000,woz2000,bra2008}. Therefore, the baseline magnitude of the DIA sources can be fainter than the limit. For unresolved sources, the instantaneous fraction of events with impact parameter $u$ small enough to be detectable scales as $u^{2}$, though over time the rate of detectable events scales as $u$. Therefore, we weight the optical depth by min($1,u^{2}$) and the rate-weighted average duration by min($1,u$), respectively. The impact parameter moments of Equation~\ref{impact} are rate-weighted as explained in Section~\ref{sec:opticaldepth} in order to reflect the fact that observables are necessarily obtained from rate-biased samples.

The finite source effect is also taken into account in our calculations. The events which involve a source star with angular radius larger than the angular Einstein radius are not used to calculate the microlensing parameters. However, they are accounted for in the source number normalisation .In practice this modification alters our results only at the 1\% level (See Equation~\ref{finite}).

\subsection{Low-mass stars and brown dwarfs}
\label{sec:hist}

\begin{figure}
\includegraphics[width=0.5\textwidth,natwidth=800,natheight=600]{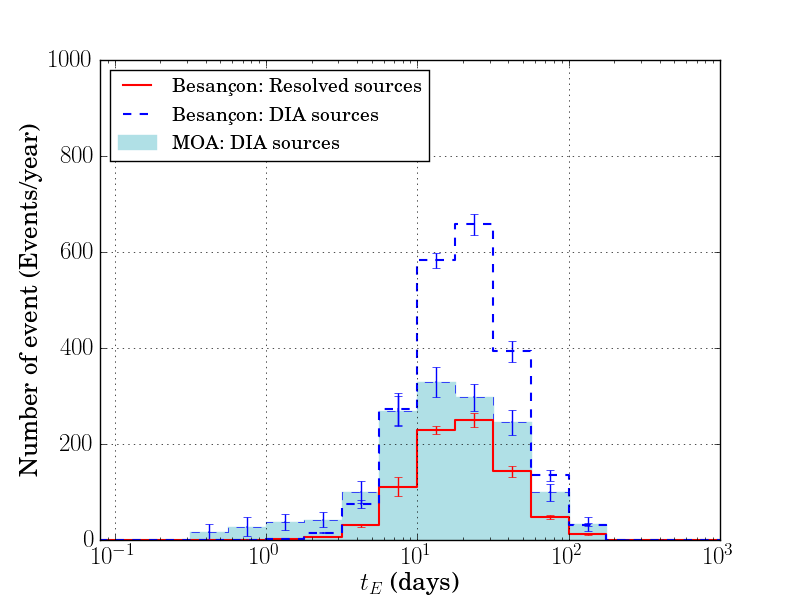}
\includegraphics[width=0.5\textwidth,natwidth=800,natheight=600]{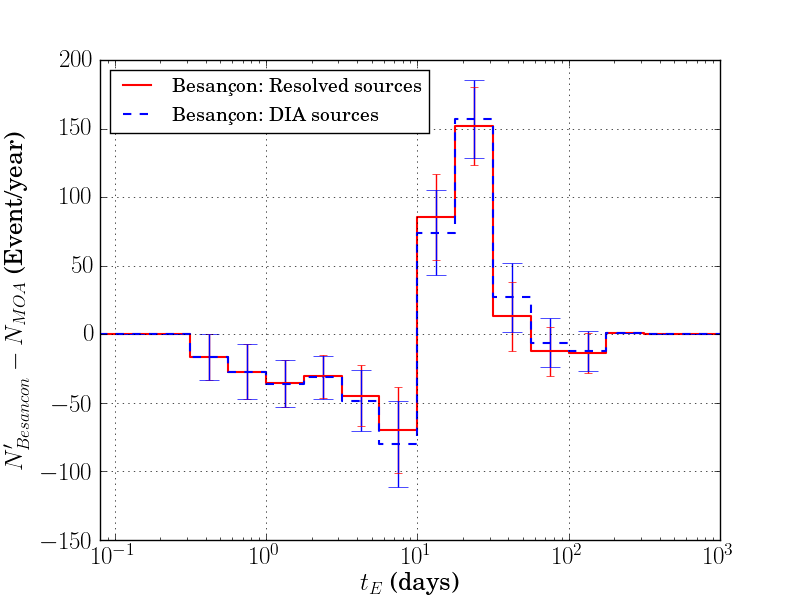}
\caption[The Besan\c{c}on Einstein radius crossing time distribution.]{The Einstein radius crossing time distribution of the MOA-II survey and the Besan\c{c}on data (top) and the scaled residual between the MOA-II survey and the Besan\c{c}on data with the MOA-II distribution error (bottom). The blue shaded area (blue thin line) represents the efficiency corrected time scale distribution for the MOA-II DIA sources, exculding event gb21-R-8-53601 (See text). The crossing time distribution of the Besan\c{c}on resolved sources (red thick line) and DIA sources (blue thick dashed line) are also presented. The error bars of Besan\c{c}on distributions are shown at 100 times their true size. For the residual, the red line and blue dashed line represent the residual of the Besan\c{c}on resolved sources and DIA sources, respectively.}
\label{fig:teHist}
\end{figure}

The time scale distribution of the MOA-II observational data, excluding the gb21-R-8-53601 event, which is located outside the Besan\c{c}on extinction map, and the Besan\c{c}on simulated data are shown in Figure~\ref{fig:teHist}. The histogram of the Besan\c{c}on data is generated from the sample catalogues using the same criteria as Section~\ref{sec:sim} but with smaller solid angles, $\Omega_{\textup{sample}}$, which contain $\sim$1,000 stars at Baade's Window in each catalogue (Table~\ref{catalogue}). From the histogram, the mean crossing time of the Besan\c{c}on resolved source (25.5~days) and DIA source (26.3~days) samples are larger than the MOA-II mean time scale for all sources (24.0 days) and RGC sources (19.2 days) \citep{sum2013}.

In order to investigate the shape of distributions, the residual event rate distribution ($N'_{Besancon} - N_{MOA}$) is shown, where $N'_{Besancon}$ is Besan\c{c}on event rate scaled to number of MOA-II events per year ($\sum N_{MOA}$). The Besan\c{c}on data shows a deficit of short time scale events ($<$10 days) and an excess of 10-30 day events which may be caused by the lack of low-mass stars and brown dwarfs in the model \citep{pen2013}. Therefore, we add in low-mass stars and brown dwarf lenses using the same stellar catalogue by replacing the lens mass according to their mass function, as discussed below. 

\begin{table}
\caption{The mass function of the simulated low-mass star population}
\centering
{\footnotesize 
\begin{tabular}{ccc}
\\
\hline\hline
\textbf{Component} & \textbf{Mass range} & \textbf{MF slope} \\ [3pt]
\hline
Thick disk & $0.079 M_{\odot}$ - $0.154 M_{\odot}$ & -0.50\\[1pt]
Bulge & $0.079 M_{\odot}$ - $0.150 M_{\odot}$ & -1.50\\[1pt]
Halo & $0.079 M_{\odot}$ - $0.085 M_{\odot}$ & -0.50\\[1pt]
\hline
\end{tabular}
}
\label{BD}
\end{table}

In each Galactic component, we add low-mass stars which are missing from the catalogue by extending the normal star mass function slopes, $\alpha \propto \log(\textup{d}N/\textup{d}M)$, to the H-burning limit of $0.079 M_{\odot}$ (Table~\ref{BD}). We also add in a brown dwarf mass function slope, $\alpha_{\textup{BD}}$, normalised to the stellar mass function at the H-burning limit and extended down to $0.001 M_{\odot}$. The added populations use the same kinetic parameters as the original catalogue and are used for the lens stars only.

\begin{figure}
\includegraphics[width=0.5\textwidth,natwidth=800,natheight=600]{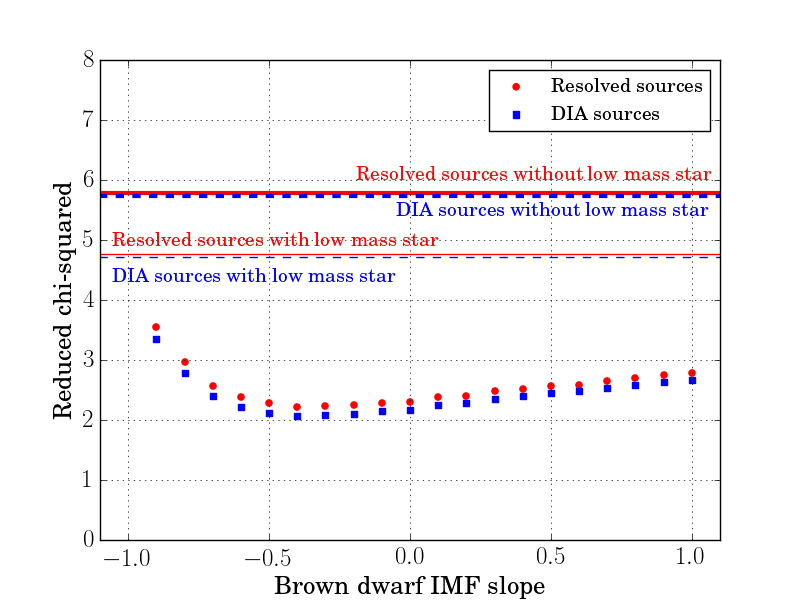}
\caption[The reduced chi-square values of the brown dwarf mass function slope.]{The reduced chi-square values of the model timescale distribution with respect to the MOA-II data for 20 fields is presented as a function of brown dwarf mass function slope. Red circle dots and lines present the Besan\c{c}on resolved source data. Blue square dots and dashed lines present the Besan\c{c}on DIA source data. Thick lines and thin line show the original data and the data with adding only low-mass stars, respectively.}
\label{fig:rechibd}
\end{figure}
In order to find the best value of $\alpha_{\textup{BD}}$, the Besan\c{c}on data from sub-field~7-4 of 20 separate fields (Fields gb1-gb20) are normalised by the MOA-II event per year and are used to calculate the timescale distributions. In Figure~\ref{fig:rechibd}, the reduced chi-squares of the predicted versus observed timescale distributions as a function of $\alpha_{\textup{BD}}$ for values of $\alpha_{\textup{BD}}$ between -0.9 and 1.0 are shown, along with reduced chi-squares of the original simulation (without adding low-mass stars and brown dwarfs) and a simulation adding only low-mass stars. The result shows that adding low-mass stars and brown dwarfs provides a better match to the MOA-II time scale distribution. \cite{sum2011} find a favoured mass function index in the brown dwarf regime, $0.01 M_{\odot}\leq M\leq 0.08M_{\odot}$, for the 2006-2007 MOA-II data is $\alpha_{\rm BD} = -0.49$. From our simulation, an MF slope of $\alpha_{\rm BD} = -0.4$ provides the best reduced chi-square value. This result is consistent with MOA-II results, but disagrees with the result from some field surveys for young brown dwarfs which suggest a power law MF with slope $\alpha_{bd}>0.0$ \citep{kir2012,jef2012}. In the following, brown dwarfs with mass function slope $-0.4$ are added in the simulation in order to bring agreement with the observed time scale distribution. 

\begin{figure}
\includegraphics[width=0.5\textwidth,natwidth=800,natheight=600]{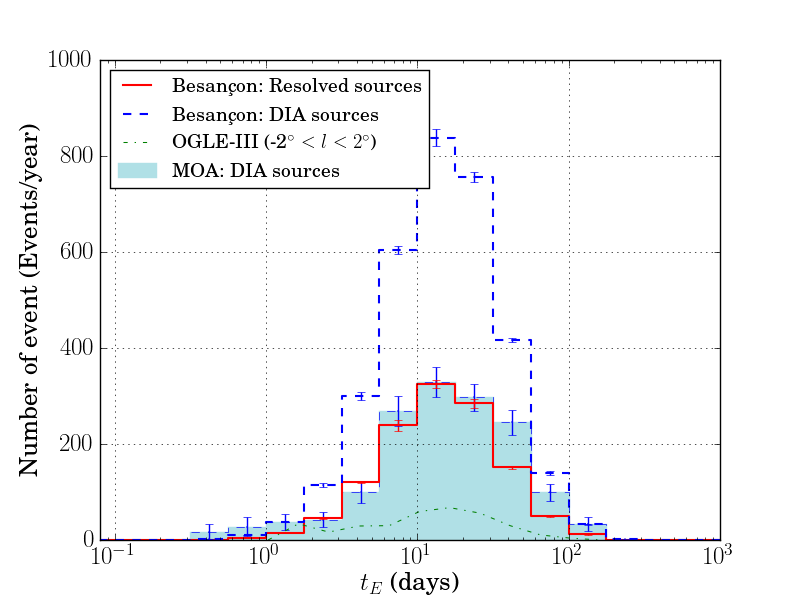}
\includegraphics[width=0.5\textwidth,natwidth=800,natheight=600]{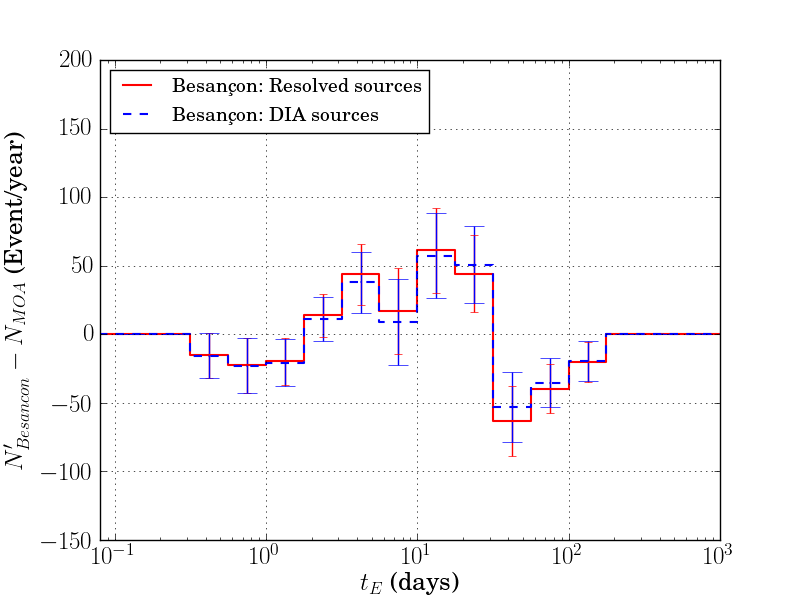}
\caption[The Besan\c{c}on timescale distribution with added low-mass stars and brown dwarfs.]{The Einstein radius crossing time distribution of the MOA-II survey, OGLE-III events in $-2^{\circ}<l<2^{\circ}$ fields and the Besan\c{c}on data with added low-mass stars and brown dwarfs (top) and the scaled residual between the MOA-II survey and the Besan\c{c}on rates (bottom). The descriptions are the same as in Figure~\ref{fig:teHist}.}
\label{fig:teHistBd}
\end{figure}
The time scale distribution of the Besan\c{c}on model with added low-mass stars is shown in Figure~\ref{fig:teHistBd}. The MOA-II survey is analysed using DIA photometry. The number of detected microlensing events per year with efficiency correction from the MOA-II survey ($N_{MOA}$) is between the number of events from the Besan\c{c}on resolved sources ($0.83\, N_{MOA}$) and DIA sources ($2.17\, N_{MOA}$). In the absence of significant blending effects, we should expect our resolved and DIA predictions to bracket the true result; the fact that it does is rather reassuring. However the effects of blending are complex and a more detailed comparison would require modeling both the source selection criteria and the source blend characteristics of the MOA-II image data. This is beyond the scope of the current paper. In the case that all resolved source events are detected, we might be tempted to conclude that 12\% of faint stars which can only be detected by the DIA method are observed. However, differences in the assumed filter response can equally be a factor. 

The mean crossing times are shorter for both resolved sources and DIA sources, at 20.3 and 20.9 days, respectively. This is close to the MOA-II RCG timescale (19.2 days), but a little lower than their mean timescale for all sources (24.0 days). These mean crossing times also compatible with mean crossing time of OGLE-III survey of resolved sources are brighter than $I<19$ mag and the relative errors on crossing time are less than 100\% with log-normal model in all three regions: positive longitude ($l>2^\circ$, 22.0 days), central ($-2^\circ<l<2^\circ$, 20.5 days) and negative longitude ($l<-2^\circ$, 24.2 days) (Figure~\ref{fig:teHistBd}) \citep{wyr2015}.

The residuals of the distribution ($\mbox{model} - \mbox{data}$) with adding low-mass stars show a slight deficit of events with short crossing time between 0.3 and 2 days and very long crossing time between 30 and 200 days. Moreover, the model tends to over-predict the number of events with duration between 2 and 30 days, though there is not a high statistical significance to any of these discrepancies. Overall, our best-fit brown dwarf slope provides a match to the MOA-II timescale distribution with a reduced $\chi^2 \simeq 2.2$ (Figure~\ref{fig:rechibd}). Therefore, the overall mass of lens population in this simulation is increased about 10\%. The total mass of each population is, thin disk $8.0\times10^{7} M_{\odot}$, thick disk $2.8\times10^{8} M_{\odot}$, halo $4.1\times10^{8} M_{\odot}$ and bulge $5.1\times10^{8} M_{\odot}$.

\begin{figure}
\includegraphics[width=0.5\textwidth,natwidth=800,natheight=600]{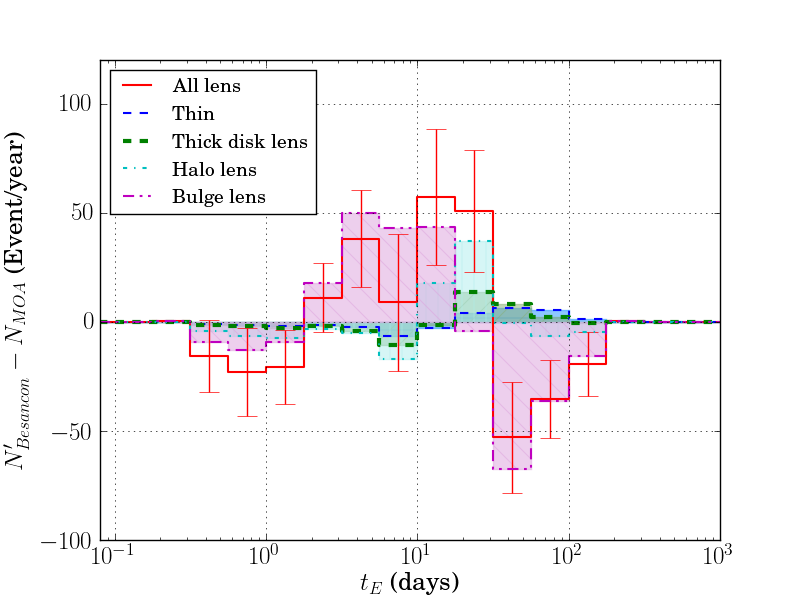}
\caption[The Einstein radius crossing time residual histogram of the Besan\c{c}on DIA sources of each galactic lens components.]{The Einstein radius crossing time residual histogram of the Besan\c{c}on DIA sources for each Galactic lens component. The residual distribution of all lens components is shown in red line. The blue thin dashed, green thick dashed, cyan dash-dotted and magenta dashed-dot-dotted lines with shaded areas represent thin disk, thick disk, halo and bulge lenses, respectively.}
\label{fig:teHistBdCom}
\end{figure}
To analyse the structure of the residual histogram, we show the contributions of each lens component separately in Figure~\ref{fig:teHistBdCom} for DIA sources. The histogram is calculated by assuming that the proportion of each component to the observed rate scales with their proportionate rate within the model. We find that the bulge lens component dominates both the over-predicted and under-predicted regimes, suggesting a mismatch in bulge kinematics, or spatial distribution, as the principal source of the problem.

\subsection{Timescale selection} \label{times}

To compare the model optical depth, rate and average duration to the MOA-II survey we must ensure that we match the timescale selection. Accordingly, the maximum Einstein crossing time ($t_{E,\textup{max}}$) of this work is set at 200 days to match the MOA-II maximum Einstein crossing time \citep{sum2013}. Furthermore, for the minimum Einstein crossing time ($t_{E,\textup{min}}$), events with duration below 40~minutes in fields gb5 and gb9 and 200~minutes in other field contribute negligibly \citep{sum2011,sum2013}. Therefore, the optical depth of all events ($\tau_{\textup{all}}$) and histograms of Einstein crossing time in each field are used to calculate optical depth ($\tau_{\textup{select}}$), average Einstein crossing time ($\left \langle t_E \right \rangle_{\textup{select}}$) and event rate ($\Gamma_{\textup{select}}$) of the events which meet the criteria. Thus
\begin{equation}
\tau_{\textup{select}}=\tau_{\textup{all}}\frac{\textstyle \sum_{i=t_{E,\textup{min}}}^{t_{E,\textup{max}}}t_{E,i}^2N_{i}}{\textstyle \sum_{i=0}^{\infty}t_{E,i}^2N_{i}} \ ,
\end{equation}
\begin{equation}
\left \langle t_{E} \right \rangle_{\textup{select}}=\frac{\textstyle \sum_{i=t_{E,\textup{min}}}^{t_{E,\textup{max}}}t_{E,i}^2N_{i}}{\textstyle \sum_{i=t_{E,\textup{min}}}^{t_{E,\textup{max}}}t_{E,i}N_{i}} \ ,
\end{equation}
and
\begin{equation}
\Gamma_{\textup{select}}=\sum_{i=t_{E,\textup{min}}}^{t_{E,\textup{max}}}N_{i} \ ,
\end{equation}
where $t_{E,i}$ and $N_{i}$ are the crossing time and the number of microlensing events associated with the {\em logarithmic}\/ timescale bin $i$, respectively.

\section{Microlensing maps}

\label{sec:besancon}

\subsection{Optical depth}
\label{sec:opticaldepth}

\begin{figure*}
{\centering
\begin{tabular}{C{0.45\textwidth}C{0.52\textwidth}}
%\begin{tabular}{lr}
\textbf{Resolved sources} & \textbf{DIA sources} \\
\end{tabular}
\subfigure{\includegraphics[width=0.50\textwidth]{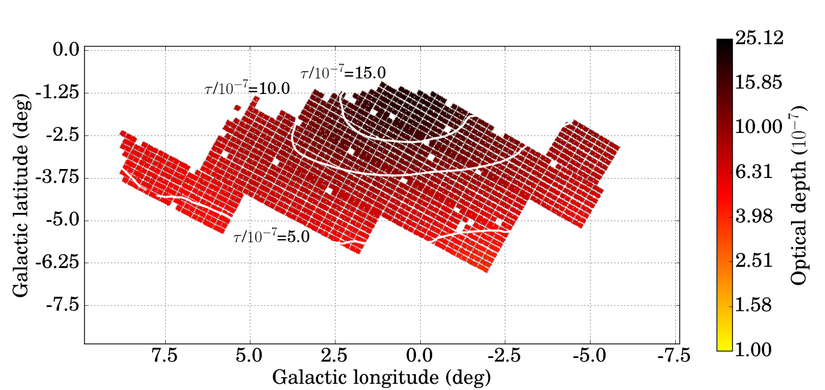}}\subfigure{\includegraphics[width=0.50\textwidth]{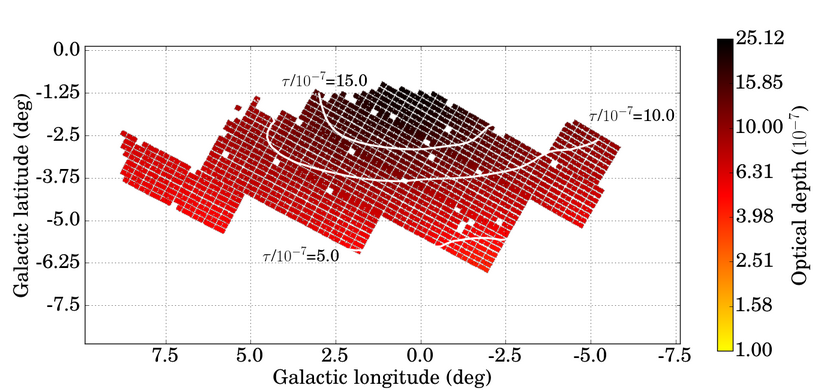}} 
(a) Optical depth\\
\subfigure{\includegraphics[width=0.50\textwidth]{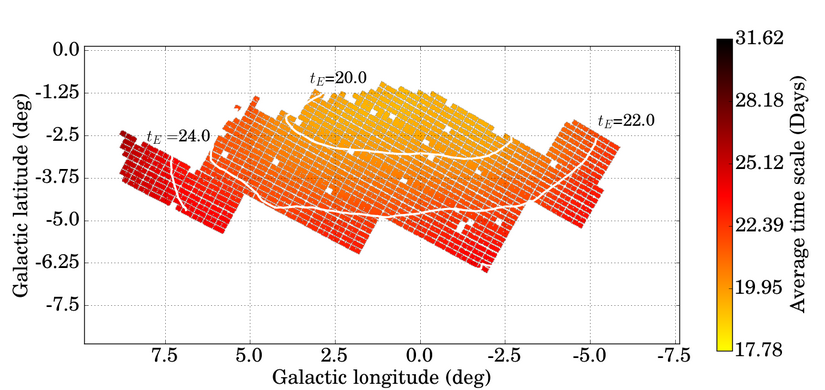}}\subfigure{\includegraphics[width=0.50\textwidth]{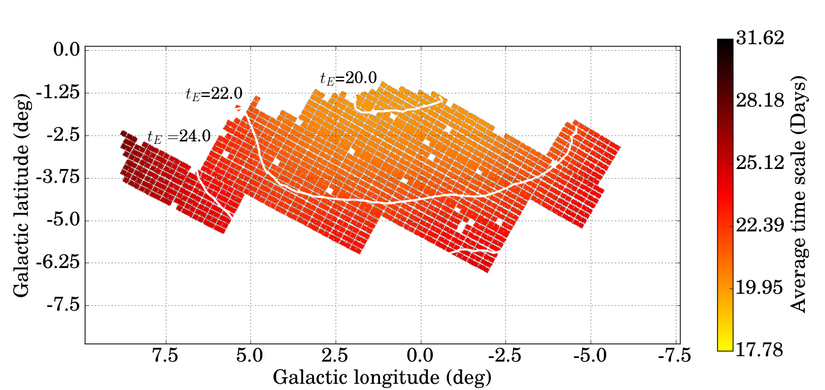}} 
(b) Average time scale\\
\subfigure{\includegraphics[width=0.50\textwidth]{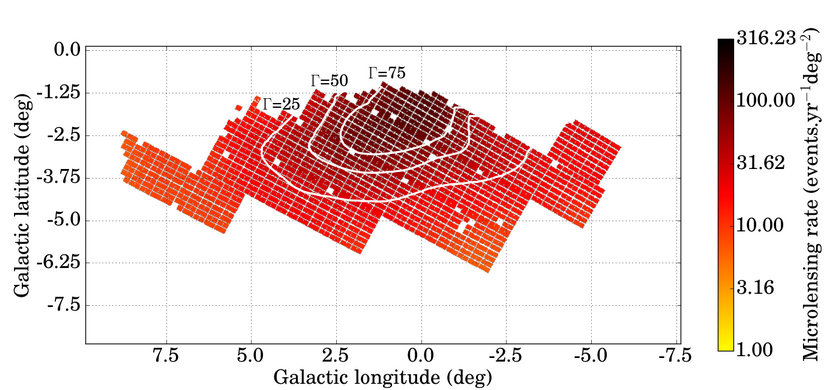}}\subfigure{\includegraphics[width=0.50\textwidth]{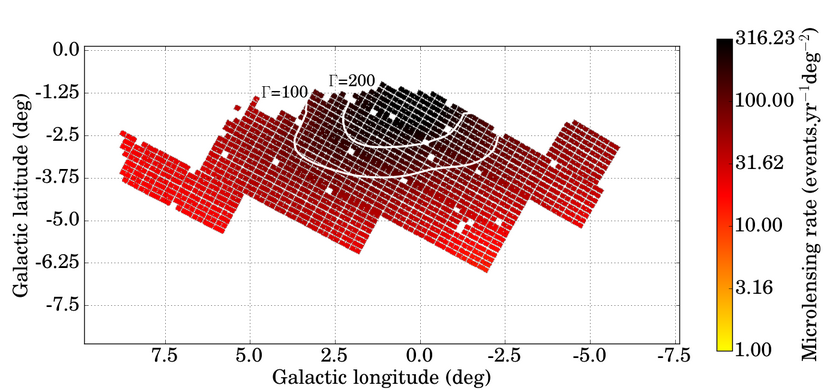}} 
(c) Microlensing event rate per square degree\\
\subfigure{\includegraphics[width=0.50\textwidth]{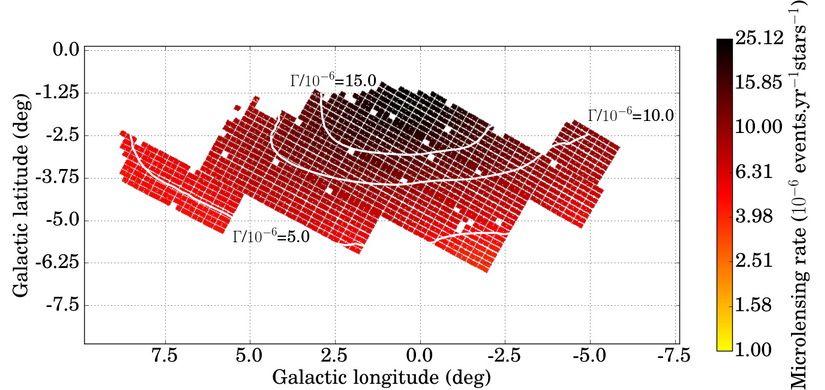}}\subfigure{\includegraphics[width=0.50\textwidth]{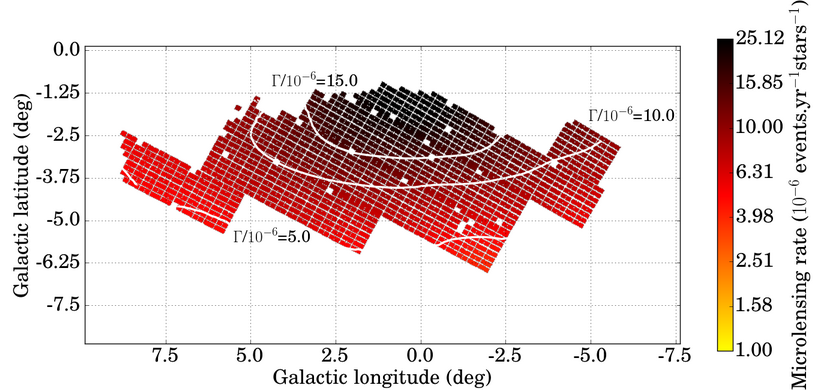}} 
(d) Microlensing event rate per star\\
}
\caption[The optical depth, average time scale, microlensing event rate per square degree and microlensing event rate per star maps]{The optical depth (a), average time scale (b), microlensing event rate per square degree (c) and microlensing event rate per star (d) for resolved sources (left) and DIA sources (right) from the Besan\c{c}on Galactic model. The maps are smoothed by the same kernel function as \cite{sum2013}. The sub-fields with $l>9^{\circ}$ are excluded.}
\label{fig:map}
\end{figure*}
Figure~\ref{fig:map}(a) shows optical depth maps for both resolved and DIA source samples for a survey limit $M_{\textup{lim}}=20$. The maps are computed for the Johnson {\it R} and {\it I} filter bands, which are comparable to the Cousins {\it R} and {\it I} bands of the MOA-II survey. The total optical depth of all source and lens pairs is calculated by averaging the optical depth of all sources along the line of sight,
\begin{equation}
\tau= \left\{ \begin{array}{ll}
 \frac{\textstyle \sum_{s=1}^{N_{s}}\sum_{l=1}^{N_{l}(D_{s}>D_{l},M_{s}<M_{\textup{lim}},u_{p} > 0)}\frac{\textstyle \pi\theta_{E}^{2}}{\textstyle \Omega_{l}}\frac{\textstyle \Omega_{0}}{\textstyle \Omega_{s}}}{\sum_{s=1}^{N_{s}(M_{s}<M_{\textup{lim}})}\frac{\textstyle \Omega_{0}}{\textstyle \Omega_{s}}} & \textup{Resolved} \ , \\
 \frac{\textstyle \sum_{s=1}^{N_{s}}\sum_{l=1}^{N_{l}(D_{s}>D_{l})}u_{p}^{2}\frac{\textstyle \pi\theta_{E}^{2}}{\textstyle \Omega_{l}}\frac{\textstyle \Omega_{0}}{\textstyle \Omega_{s}}}{\textstyle \sum_{s=1}^{N_{s}}\left \langle u_{p}^{2} \right \rangle_{w}\frac{\textstyle \Omega_{0}}{\textstyle \Omega_{s}}} & \textup{DIA} \ , \\
\end{array} \right.
\end{equation}
where $M_s$ is the source magnitude and $M_{\rm lim}$ is the survey limiting magnitude. $\Omega_{0}$ is the MOA-II sub-field solid angle. $\Omega_{l}$ and $\Omega_{s}$ are the solid angles over which the source and lens catalogues are simulated, respectively, and $N_{s}$ and $N_{l}$ are number of catalogue sources and lenses, respectively. The impact parameter $u_p$ is given by
\begin{equation}
u_{p}= \left\{ \begin{array}{ll}
 0, & \theta_{E}\times \textup{min}(1,u_{t}) \leq \theta_{*} \ , \\
 \textup{min}(1,u_{t}), & \textup{otherwise} \ . \\
\end{array} \right.
\label{finite}
\end{equation}
Here, $\theta_{*}$ is the angular star radius and $u_{t}$ is the largest impact parameter for an event to be detectable above the survey limiting sensitivity. $D_{s}$ and $D_{l}$ are the distance to the source and the lens from the observer, respectively. To take account of magnification suppression by finite source size effects, whenever under the point-source regime the source angular size is larger than the largest detectable impact parameter, we assume the event to be undetectable. The $n$-th moment of $u_p$, $\langle u_p^n \rangle_w$, is obtained through rate-weighted averaging:
\begin{equation}
\left \langle u_p^n \right \rangle_w = \left\{ \begin{array}{ll}
 1, & M_{s}<M_{\textup{lim}} \ , \\
 \frac{\textstyle \sum_{l=1}^{N_{l}(D_{s}>D_{l})}u_p^n \mu D_{l}R_{E}}{\textstyle \sum_{s=1}^{N_{s}}\sum_{l=1}^{N_{l}(D_{s}>D_{l})}\mu D_{l}R_{E}}, & \textup{otherwise} \ , \\
\end{array} \right.
\label{impact}
\end{equation}
where $\mu$ is the lens--source pair-wise relative proper motion.

We employ the same Gaussian spatial smoothing window function as \cite{sum2013}, with $\sigma$ = 0.4$^{\circ}$ within 1$^{\circ}$. In order to compare the data with the MOA-II results, the sub-fields with $l>9^{\circ}$ are excluded due to the kernel contribution from sub-fields outside the Besan\c{c}on extinction map at $l>10^{\circ}$.

\begin{figure*}
{\centering
\subfigure{\includegraphics[width=0.5\textwidth]{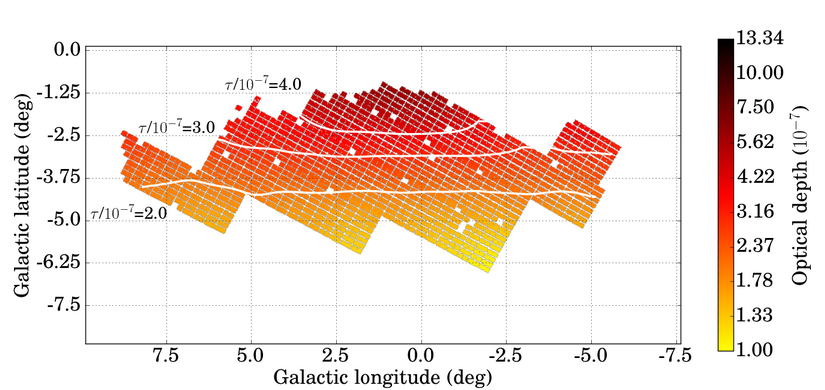}}\subfigure{\includegraphics[width=0.5\textwidth]{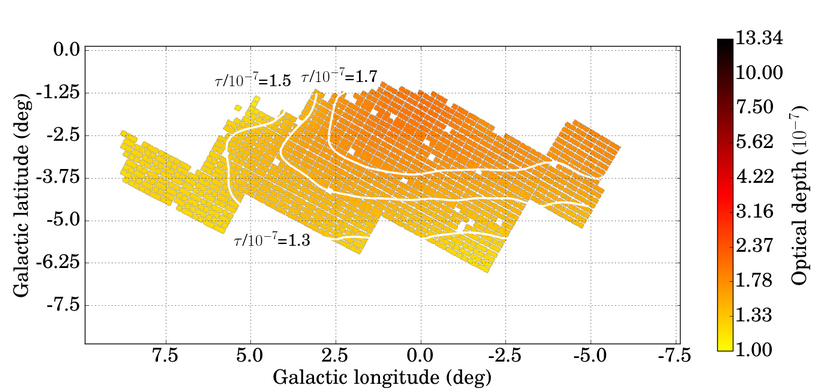}} 
\begin{tabular}{C{0.45\textwidth}C{0.5\textwidth}}
(a) Thin disk & (b) Thick disk \\
\end{tabular}
\subfigure{\includegraphics[width=0.5\textwidth]{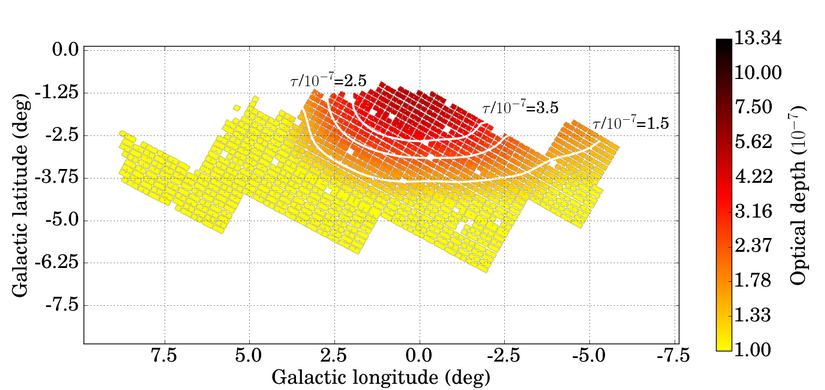}}\subfigure{\includegraphics[width=0.5\textwidth]{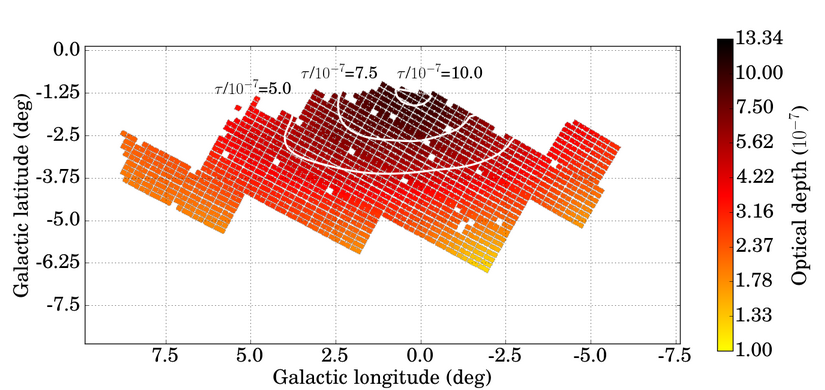}} 
\begin{tabular}{C{0.45\textwidth}C{0.5\textwidth}}
(c) Halo & (d) Bulge \\
\end{tabular}
}
\caption[The DIA optical depth maps of each Galactic lens component with the DIA method]{The optical depth maps of each Galactic lens component: (a) thin disk lenses, (b) thick disk lenses, (c) stellar halo lenses and (d) bulge lenses. The maps have same description as the map in Figure~\ref{fig:map}}
\label{fig:mapO}
\end{figure*}

From the simulation results, the optical depth of DIA sources is larger than the optical of resolved sources, as was also found by \cite{ker2009} using an earlier version of the Besa\c{c}on model. However the current model predicts a significantly lower optical depth compared with \cite{ker2009} due to the lower mass of the Galactic bulge (which is a factor of two lower than for the earlier model). However, the optical depth values are compatible with the \cite{pen2013} result which also uses a more recent version of the Besan\c{c}on model (version 1106). In Figure~\ref{fig:mapO}, the optical depth distribution is dominated by the bulge population which contains about 50-80\% of the stars. The thin disk, thick disk and stellar halo lenses provide slightly larger optical depth contributions at negative longitude than positive longitude due to the fact that bulge sources, which dominate the statistics, tend to lie at larger distances at negative longitudes. 

\subsection{Maps of average event duration}

In order to calculate the average time scale, each Einstein crossing time ($t_{E}$) is rate-weighted by a factor $\mu D_{l}R_{E}$. Finally, the average time scale, $\left \langle t_E \right \rangle$, is obtained as
\begin{equation}
\left \langle t_E \right \rangle= \left\{ \begin{array}{l}
 \frac{\textstyle \sum_{s=1}^{N_{s}}\sum_{l=1}^{N_{l}(D_{s}>D_{l},M_{s}<M_{\textup{lim}},u_{p} >0)}\theta_{E}^{2}D_{l}^{2}\frac{\textstyle \Omega_{0}}{\textstyle \Omega_{l}\Omega_{s}}}{\textstyle \sum_{s=1}^{N_{s}}\sum_{l=1}^{N_{l}(D_{s}>D_{l},M_{s}<M_{\textup{lim}})}\theta_{E}D_{l}^{2}\mu\frac{\textstyle \Omega_{0}}{\textstyle \Omega_{l}\Omega_{s}}} \\
 \quad\quad \quad\quad \quad\quad \quad\quad \quad\quad \quad\quad \quad\quad \quad\quad \textup{(Resolved)}, \\
 \\
 \frac{\textstyle \sum_{s=1}^{N_{s}}\sum_{l=1}^{N_{l}(D_{s}>D_{l})}u_{p}\theta_{E}^{2}D_{l}^{2}\frac{\textstyle \Omega_{0}}{\textstyle \Omega_{l}\Omega_{s}}}{\textstyle \sum_{s=1}^{N_{s}}\sum_{l=1}^{N_{l}(D_{s}>D_{l})}\left \langle u_{p} \right \rangle_{w}\theta_{E}D_{l}^{2}\mu\frac{\textstyle \Omega_{0}}{\textstyle \Omega_{l}\Omega_{s}}} \\
 \quad\quad \quad\quad \quad\quad \quad\quad \quad\quad \quad\quad \quad\quad \quad\quad \textup{(DIA)}, \\
\end{array} \right.
\end{equation}

Maps of the average event duration are shown in Figure~\ref{fig:map}(b). The maps show shorter timescales compared to \cite{ker2009} and \cite{pen2013}, due to the addition of low-mass star and brown dwarf lenses. There is no major difference between the average time scale of resolved sources and DIA sources. The negative longitudes provide slightly longer time scales than positive longitudes due to the bar geometry resulting in typically larger Einstein radii at negative longitudes.

\begin{figure*}
{\centering
\subfigure{\includegraphics[width=0.5\textwidth]{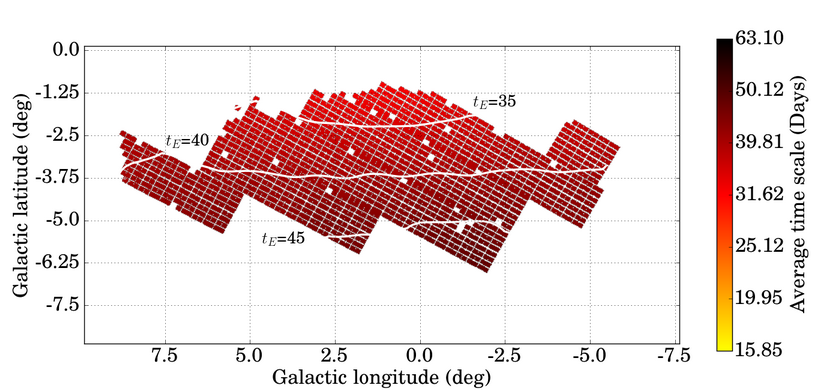}}\subfigure{\includegraphics[width=0.5\textwidth]{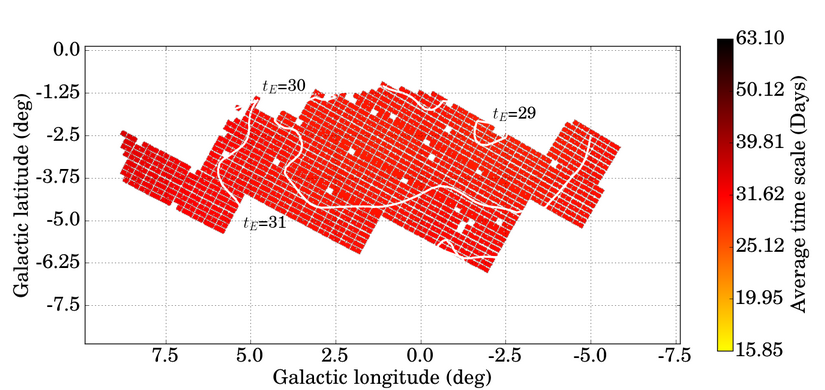}} 
\begin{tabular}{C{0.45\textwidth}C{0.5\textwidth}}
(a) Thin disk & (b) Thick disk \\
\end{tabular}
\subfigure{\includegraphics[width=0.5\textwidth]{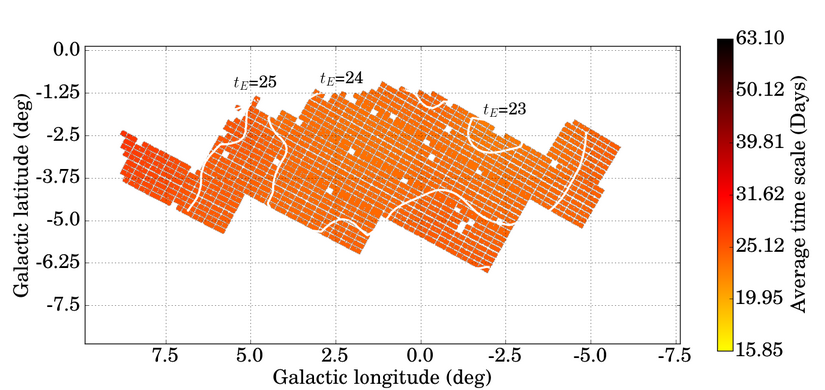}}\subfigure{\includegraphics[width=0.5\textwidth]{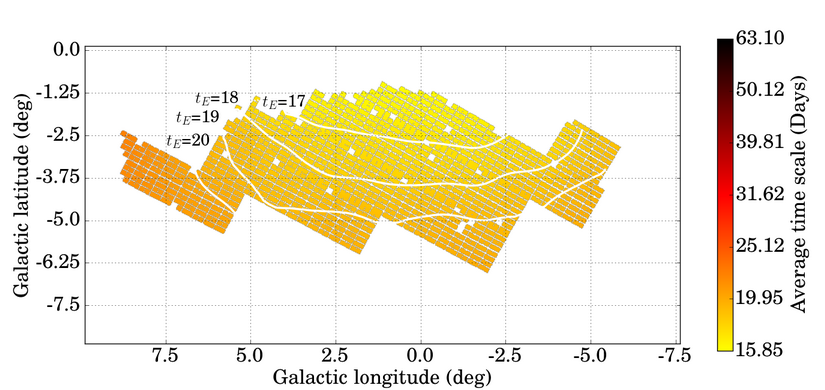}} 
\begin{tabular}{C{0.45\textwidth}C{0.5\textwidth}}
(c) Halo & (d) Bulge \\
\end{tabular}
}
\caption[The average time scale maps of each Galactic lens component for DIA sources]{The average time scale maps of each Galactic lens component for DIA sources: (a) thin disk, (b) thick disk, (c) halo and (d) bulge. The maps have same description as the map in Figure~\ref{fig:map}}
\label{fig:mapT}
\end{figure*}

In Figure~\ref{fig:mapT}, we show the average timescale maps individually for the thin disk, thick disk, stellar halo lens and bulge lens populations. The maps show a reasonably symmetric spatial distribution in the average event duration, with bulge lenses exhibiting typically shorter time scales compared to the other lens components. Since bulge lenses dominate the event rate in the inner Galaxy (Figure~\ref{fig:mapE}) the overall map of event duration shown in Figure~\ref{fig:map}(b) closely resembles that of the bulge lens population. We also confirm from Figure~\ref{fig:mapT} that the long duration region at longitude $l > 7.5^{\circ}$ evident in Figure~\ref{fig:map}(b) arises from the disk lens population as the density of bulge lenses become sub-dominant away from the Galactic Centre.

\subsection{Map of microlensing event rate}

The total event rate is obtained simply by dividing the optical depth maps by their corresponding average time scale maps:
\begin{equation}
\Gamma=\frac{2}{\pi}\frac{\tau}{\left \langle t_{E} \right \rangle} \ .
\end{equation}

Figure~\ref{fig:map}(c) and Figure~\ref{fig:map}(d) show maps of microlensing event rate per square degree ($\Gamma_{\textup{deg}^{2}}$) and event rate per star ($\Gamma_{\textup{star}}$), respectively. $\Gamma_{\textup{deg}^{2}}$ is obtained by integrating the rate over the effective number of sources:
\begin{equation}
N= \left\{ \begin{array}{ll}
 \sum_{s=1}^{N_{s}(M_{s}>M_{\textup{lim}})}\frac{\textstyle \Omega_{0}}{\textstyle \Omega_{s}} & \textup{Resolved} \ , \\
 \textstyle \sum_{s=1}^{N_{s}}\left \langle u_{p} \right \rangle_{w}\frac{\textstyle \Omega_{0}}{\textstyle \Omega_{s}} & \textup{DIA} . \\
\end{array} \right.
\end{equation}

In Figure~\ref{fig:map}(c) we see that $\Gamma_{\textup{deg}^{2}}$ for DIA sources is higher than for resolved sources, as expected. The area integrated microlensing event rate in the simulated maps for resolved sources and DIA sources is 1,250 and 3,250 events per year, respectively. The maps of $\Gamma_{\textup{star}}$ in Figure~\ref{fig:map}(d) for resolved sources and DIA sources do not show a major difference indicating that, overall, they probe sources and lenses at similar distances with similar kinematics.

\begin{figure*}
{\centering
\subfigure{\includegraphics[width=0.5\textwidth]{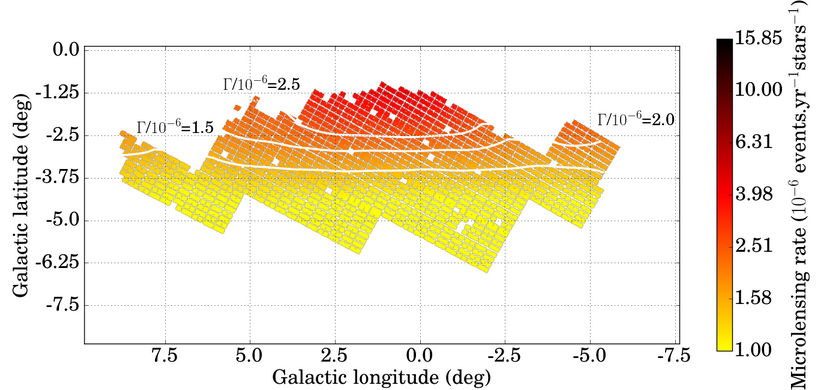}}\subfigure{\includegraphics[width=0.5\textwidth]{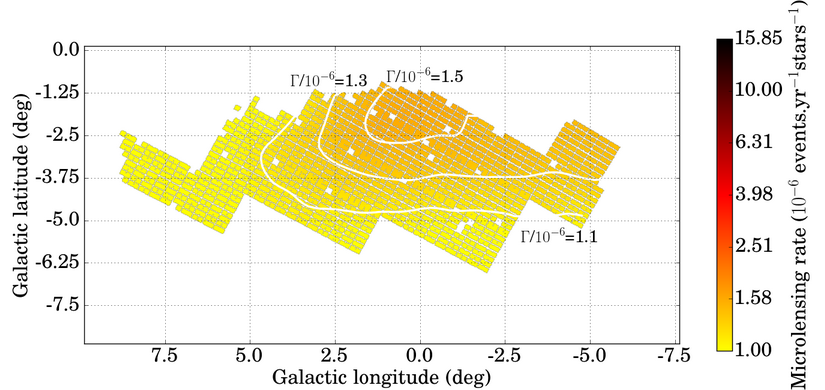}}
\begin{tabular}{C{0.45\textwidth}C{0.5\textwidth}}
(a) Thin disk & (b) Thick disk \\
\end{tabular}
\subfigure{\includegraphics[width=0.5\textwidth]{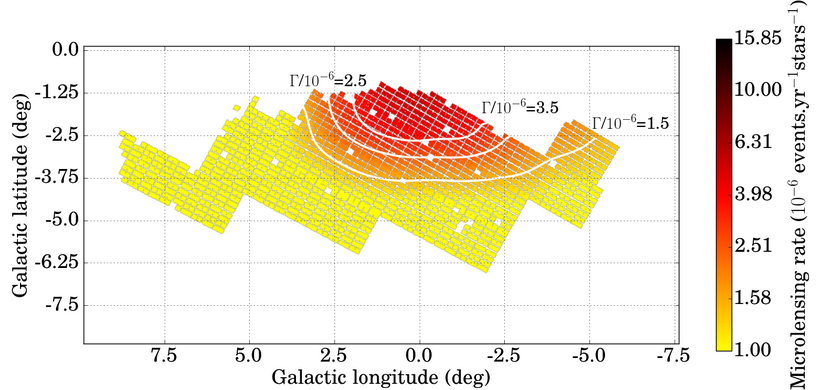}}\subfigure{\includegraphics[width=0.5\textwidth]{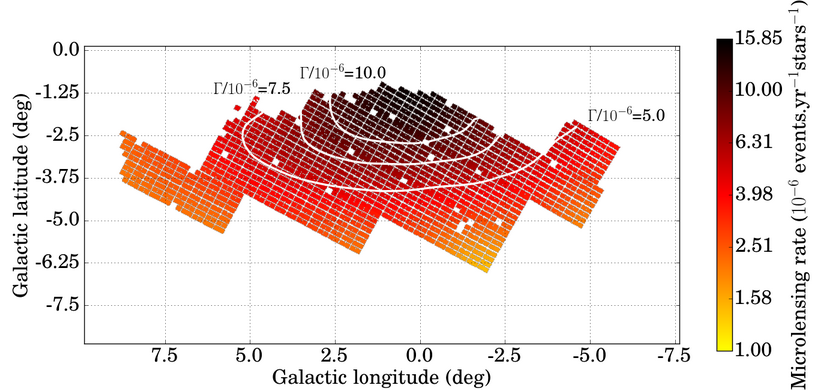}}
\begin{tabular}{C{0.45\textwidth}C{0.5\textwidth}}
(c) Halo & (d) Bulge \\
\end{tabular}
}
\caption[The microlensing event rate per star maps of each Galactic lens component for DIA sources]{The microlensing event rate per star maps of each Galactic lens component for DIA sources: (a) thin disk, (b) thick disk, (c) halo and (d) bulge. The maps have same description as the map in Figure~\ref{fig:map}}
\label{fig:mapE}
\end{figure*}

In Figure~\ref{fig:mapE}, the maps of $\Gamma_{\textup{star}}$ are shown separately for each lens population. The strong dominance of bulge lenses over most of the MOA-II region is evident.

\section{Confronting MOA-II observations}

\subsection{Galactic latitude variation}
\label{sec:opfit}

\begin{figure*}
{\centering
\includegraphics[width=0.52\textwidth,natwidth=800,natheight=600]{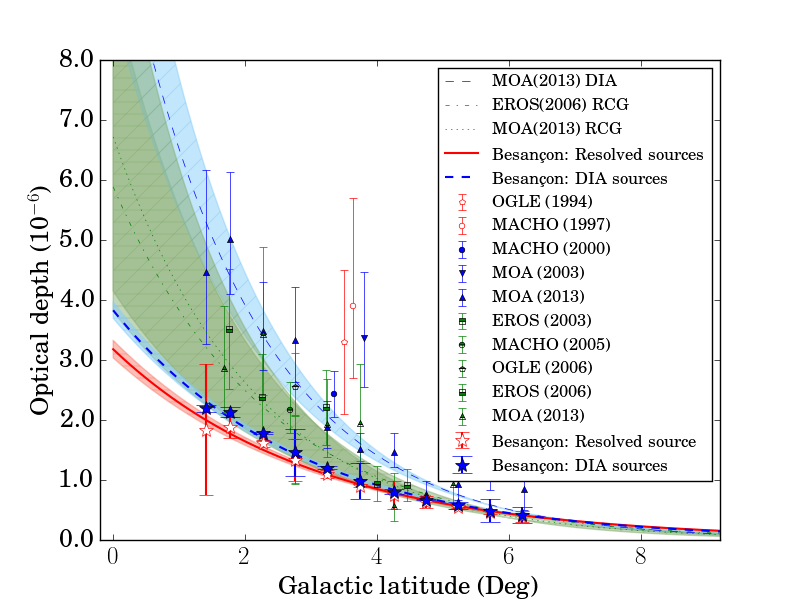} \\
\includegraphics[width=0.52\textwidth,natwidth=800,natheight=600]{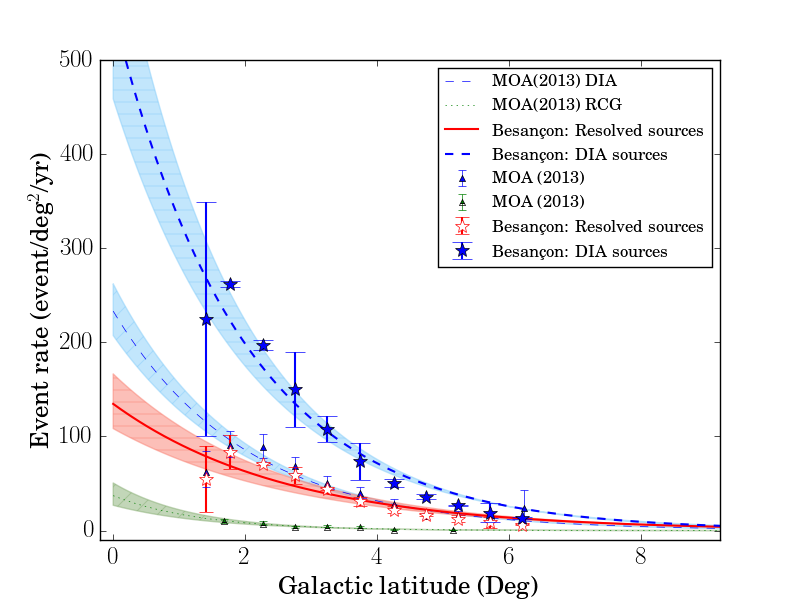} \\
\includegraphics[width=0.52\textwidth,natwidth=800,natheight=600]{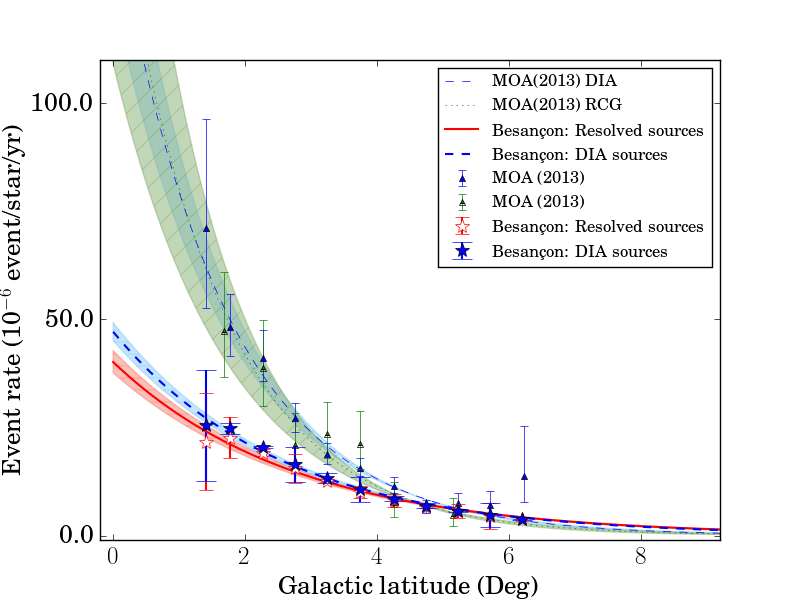}
}
\caption[Optical depth and microlensing event rate as a function of Galactic latitude.]{The optical depth (top), microlensing event rate per square degree (middle) and microlensing event rate per star (bottom) as a function of Galactic latitude. The measurements are averaged over Galactic longitudes $-5^{\circ} < l < 5^{\circ}$. Different markers represent different survey measurements (See Table~\ref{OMlist}): OGLE (pentagon), MACHO (circle), MOA (triangle), EROS (square) and simulated data from the Besan\c{c}on galactic model (star) (See Table~\ref{OMlistB}). Results of resolved sources, DIA sources and RCG source are presented with unfilled, filled and half-filled makers. The error bars of the Besan\c{c}on simulation results are shown at 100 times their true size. The thin dashed, dash-dotted and dotted lines represent fits to the MOA-II all-source sample, EROS RCG sample and MOA-II RCG sample, respectively \citep{ham2006,sum2013}. The thick solid and dashed lines are fits to the resolved source and DIA source simulations of this paper. The shaded areas represent the 68\% confidence interval of EROS, MOA-II and Besan\c{c}on fits, respectively (See text).}
\label{fig:opfit}
\end{figure*}

The optical depth, microlensing event rate per square degree and microlensing event rate per star from the Besan\c{c}on Galactic model and survey observations at different Galactic latitudes are presented in Figure~\ref{fig:opfit}. The results are calculated from the optical depth and microlensing event rate between $l=-5^{\circ}$ and $l=5^{\circ}$. The stars in each simulated sub-field are binned to 0.5$^{\circ}$ in Galactic latitude, in similar fashion to the MOA-II survey \cite{sum2013}. The results from the previous measurements, as well as the simulated models of the present paper, are listed in Table~\ref{OMlist} and~\ref{OMlistB}, respectively. The shaded areas in Figure~\ref{fig:opfit} show 68\% confidence interval of the data. The shaded $68\%$ confidence intervals are obtained by generating random deviate distributions of each exponential fit model assuming that errors on the best-fit parameters are Gaussian distributed.

From the MOA-II data and the Besan\c{c}on simulated data, the optical depth at $b>1.5^{\circ}$ agrees very well with an exponential fit. For $b<1.8^{\circ}$, the optical depths decrease due to the high column density of dust in that area. Over lower latitude regions ($b<3^{\circ}$), the Besan\c{c}on DIA optical depth is lower than the MOA-II all-source optical depth by a factor of 2, a factor similar to that found by \cite{pen2013}. The exponential models of the Besan\c{c}on optical depth are,
\begin{equation}
\tau_{\textup{Res}} = (1.18\pm0.03)\, e^{(0.330\pm0.017)(3-\left | b \right |)}\times10^{-6} \ , \\ \nonumber
\end{equation}
\begin{equation}
\tau_{\textup{DIA}} = (1.31\pm0.02)\, e^{(0.357\pm0.013)(3-\left | b \right |)}\times10^{-6} \ . \\
\end{equation}

The event rate per square degree for the Besan\c{c}on resolved sources is compatible with the MOA-II all-source event rate, however for DIA sources it is 3 times higher than the MOA-II result. This might be a consequence of the blending effect as discussed in Section~\ref{sec:hist}. The results from both also show the same turning point at $l=1.8^{\circ}$ as the optical depth. The exponential fits for the event rate per square degree give,
\begin{equation}
\Gamma_{\textup{deg}^{2},\textup{Res}} = (43\pm4)\, e^{0.380\pm0.510)(3-\left | b \right |)} \ , \\ \nonumber
\end{equation}
\begin{equation}
\Gamma_{\textup{deg}^{2},\textup{DIA}} = (119\pm9)\, e^{(0.510\pm0.060)(3-\left | b \right |)} \ . \\
\end{equation}

For the microlensing event rate per star, the results are similar to the optical depth result, which is expected given the general agreement of the average event duration. The exponential models of the simulated event rate per star can be written as,
\begin{equation}
\Gamma_{\textup{star},\textup{Res}} = (13.5\pm0.4)\,e^{(0.362\pm0.023)(3-\left | b \right |)}\times10^{-6} \ , \\ \nonumber
\end{equation}
\begin{equation}
\Gamma_{\textup{star},\textup{DIA}} = (14.6\pm0.3)\,e^{(0.391\pm0.016)(3-\left | b \right |)}\times10^{-6} \ . \\
\end{equation}

Whilst the model average event duration is in reasonable agreement with the MOA-II observations the factor 2 discrepancy with both the optical depth and rate suggests that the model bulge mass is too low to accommodate the microlensing results. The bulge mass would need to be increased by a factor 2.6 in order to match the overall optical depth distribution. However, we note that such a change could not be made in a self-consistent manner without also altering the lens and source kinematics and the event timescale distribution. 
\begin{table*}
\caption{Observed microlensing optical depth and rate measurements towards the Galactic Centre.}
\centering
{\footnotesize 
\begin{tabular}{lcccccccc}
\\
\hline\hline
\textbf{Project} & \textbf{Field} & \textbf{Method} & $t_{E,max}$ & \textbf{$N_{event}$} & $l,b^{*}$ & $\tau$ & $\Gamma_{\textup{deg}^{2}}$& $\Gamma_{\textup{star}}$\\ [3pt]
& deg$^{2}$ & & days & events & & $\times10^{6}$ & yr$^{-1}$deg$^{-2}$ & $\times 10^{-6}$yr$^{-1}$star$^{-1}$\\ [3pt]
\hline
OGLE(1994)$^{[1]}$ & 0.81 & Resolved & 100 & 9 & $\pm$5$^{\circ}$,-3.5$^{\circ}$&3.3$\pm$1.2&-&-\\
MACHO(1997)$^{[2]}$ & 12.0 & Resolved & 150 & 45 & 2.55$^{\circ}$,3.64$^{\circ}$ & 3.9$^{+1.8}_{-1.2}$&-&-\\
MACHO(2000)$^{[3]}$ & 4.0 & DIA & 150 & 99 & 2.68$^{\circ}$,-3.35$^{\circ}$ & 2.43$^{+0.39}_{-0.38}$&-&-\\
EROS(2003)$^{[4]}$ & 15.0 & RCG & 400 & 16 & 2.5$^{\circ}$,-4.0$^{\circ}$ & 0.94$\pm$0.29&-&-\\
MOA(2003)$^{[5]}$ & 18.0 & DIA & 150 & 28 & 3.0$^{\circ}$,-3.8$^{\circ}$ & 2.59$^{+0.84}_{-0.64}$&-&-\\
MACHO(2005)$^{[6]}$ & 4.5 & RCG & 350 & 62 & 1.5$^{\circ}$,-2.68$^{\circ}$ & 2.17$^{+0.47}_{-0.38}$&-&-\\
OGLE(2006)$^{[7]}$ & 5.0 & RCG & 400 & 32 & 1.16$^{\circ}$,-2.75$^{\circ}$ & 2.55$^{+0.57}_{-0.46}$&-&-\\
EROS(2006)$^{[8]}$ & 66.0 & RCG & 400 & 25 & (-6.00$^{\circ}$-10.00$^{\circ}$), 1.75$^{\circ}$& 3.52$\pm$1.00&-&-\\
&&&&22 & (-6.00$^{\circ}$,10.00$^{\circ}$), 2.26$^{\circ}$& 2.38$\pm$0.72&-&-\\
&&&&24 & (-6.00$^{\circ}$,10.00$^{\circ}$), 2.76$^{\circ}$& 1.31$\pm$0.38&-&-\\
&&&&25 & (-6.00$^{\circ}$,10.00$^{\circ}$), 3.23$^{\circ}$& 2.21$\pm$0.62&-&-\\
&&&&24 & (-6.00$^{\circ}$,10.00$^{\circ}$), 4.45$^{\circ}$& 0.92$\pm$0.72&-&-\\
MOA(2013)$^{[9]}$ & 3.2 & DIA & 200 & 12 & (-5.00$^{\circ}$,5.00$^{\circ}$), -1.40$^{\circ}$ & 4.47$^{+1.69}_{-1.21}$& 62.4$^{+22.1}_{-16.3}$& 71.2$^{+25.2}_{-18.6}$\\
&&&&52 & (-5.00$^{\circ}$,5.00$^{\circ}$), -1.77$^{\circ}$ & 5.01$^{+1.12}_{-0.91}$& 90.9$^{+14.5}_{-12.6}$& 48.2$^{+7.7}_{-6.7}$\\
&&&&70 & (-5.00$^{\circ}$,5.00$^{\circ}$), -2.26$^{\circ}$ & 3.49$^{+0.81}_{-0.66}$& 88.6$^{+13.7}_{-11.7}$& 41.1$^{+6.4}_{-5.4}$\\
&&&&75 & (-5.00$^{\circ}$,5.00$^{\circ}$), -2.76$^{\circ}$ & 3.33$^{+0.88}_{-0.69}$& 68.8$^{+9.0}_{-7.9}$& 27.1$^{+3.5}_{-3.1}$\\
&&&&67 & (-5.00$^{\circ}$,5.00$^{\circ}$), -3.25$^{\circ}$ & 1.88$^{+0.43}_{-0.35}$& 50.6$^{+7.0}_{-6.3}$& 18.8$^{+2.6}_{-2.3}$\\
&&&&58 & (-5.00$^{\circ}$,5.00$^{\circ}$), -3.75$^{\circ}$ & 1.52$^{+0.26}_{-0.23}$& 40.3$^{+5.8}_{-5.2}$& 15.7$^{+2.2}_{-2.0}$\\
&&&&43 & (-5.00$^{\circ}$,5.00$^{\circ}$), -4.25$^{\circ}$ & 1.47$^{+0.32}_{-0.26}$& 28.6$^{+4.9}_{-4.2}$& 11.6$^{+2.0}_{-1.7}$\\
&&&&22 & (-5.00$^{\circ}$,5.00$^{\circ}$), -4.74$^{\circ}$ & 0.76$^{+0.22}_{-0.18}$& 15.0$^{+3.6}_{-2.9}$& 6.6$^{+1.6}_{-1.3}$\\
&&&&16 & (-5.00$^{\circ}$,5.00$^{\circ}$), -5.23$^{\circ}$ & 0.94$^{+0.39}_{-0.28}$& 16.2$^{+4.6}_{-3.6}$& 7.6$^{+2.2}_{-1.7}$\\
&&&&8 & (-5.00$^{\circ}$,5.00$^{\circ}$), -5.72$^{\circ}$ & 1.34$^{+0.84}_{-0.51}$& 13.4$^{+5.8}_{-4.0}$& 7.2$^{+3.1}_{-2.1}$\\
&&&&4 & (-5.00$^{\circ}$,5.00$^{\circ}$), -6.23$^{\circ}$ & 0.85$^{+0.64}_{-0.35}$& 23.7$^{+19.5}_{-10.4}$& 13.9$^{+11.5}_{-6.1}$\\
&& RCG & 200 & 16 & (-5.00$^{\circ}$,5.00$^{\circ}$), -1.69$^{\circ}$ & 2.87$^{+1.03}_{-0.75}$& 9.8$^{+2.8}_{-2.2}$& 47.3$^{+13.6}_{-10.6}$\\
&&&&16 & (-5.00$^{\circ}$,5.00$^{\circ}$), -2.26$^{\circ}$ & 3.44$^{+1.45}_{-1.03}$& 7.5$^{+2.2}_{-1.7}$& 38.7$^{+11.2}_{-8.7}$\\
&&&&11 & (-5.00$^{\circ}$,5.00$^{\circ}$), -2.76$^{\circ}$ & 1.40$^{+0.67}_{-0.45}$& 3.9$^{+1.4}_{-1.0}$& 20.9$^{+7.4}_{-5.5}$\\
&&&&14 & (-5.00$^{\circ}$,5.00$^{\circ}$), -3.25$^{\circ}$ & 1.93$^{+0.76}_{-0.55}$& 4.3$^{+1.3}_{-1.0}$& 23.6$^{+7.3}_{-5.6}$\\
&&&&11 & (-5.00$^{\circ}$,5.00$^{\circ}$), -3.75$^{\circ}$ & 1.95$^{+0.98}_{-0.64}$& 3.5$^{+1.3}_{-0.9}$& 21.2$^{+7.7}_{-5.7}$\\
&&&&4 & (-5.00$^{\circ}$,5.00$^{\circ}$), -4.25$^{\circ}$ & 0.57$^{+0.55}_{-0.26}$& 1.1$^{+0.8}_{-0.4}$& 7.3$^{+5.0}_{-2.9}$\\
&&&&3 & (-5.00$^{\circ}$,5.00$^{\circ}$), -5.15$^{\circ}$ & 0.93$^{+0.99}_{-0.41}$& 0.6$^{+0.5}_{-0.3}$& 4.7$^{+3.9}_{-2.4}$\\
\hline
\end{tabular}
}
\raggedright{$^{*}$ The values of Galactic latitude ($l$) and Galactic longitude ($b$) shown in Table~\ref{OMlist} are average position of the map or average Galactic latitude of field in each Galactic longitude bin.} \\
\raggedright{\textbf{Note} [1]: \citep{uda1994}, [2]: \citep{alc1997}, [3]: \citep{alc2000}, [4]: \citep{afo2003}, [5]: \citep{sum2003}, [6]: \citep{pop2005}, [7]: \citep{sum2006}, [8]: \citep{ham2006}, [9]: \citep{sum2013}}
\label{OMlist}
\end{table*}

\begin{table*}
\caption{The Besan\c{c}on model microlensing optical depth and event rate within $|l| < 5^{\circ}$ using $0.5^{\circ}$ bins in $b$.}
\centering
{\footnotesize 
\begin{tabular}{ccccc}
\\
\hline\hline
$\left \langle b \right \rangle$ & \textbf{$N_{event}$} & $\tau$ & $\Gamma_{\textup{deg}^{2}}$ & $\Gamma\textup{star}$\\ [3pt]
& events & $\times10^{7}$ & yr$^{-1}$deg$^{-2}$ & $\times 10^{-6}$yr$^{-1}$star$^{-1}$\\ [3pt]
\hline
\multicolumn{5}{c}{\textbf{Resolved source}} \\
\hline
-1.40$^{\circ}$ & 18.414 & 18.414 $\pm$ 0.109 & 54.828 $\pm$ 0.347 & 21.774 $\pm$ 0.111 \\
-1.77$^{\circ}$ & 18.759 & 18.759 $\pm$ 0.017 & 83.258 $\pm$ 0.183 & 22.671 $\pm$ 0.048 \\
-2.26$^{\circ}$ & 16.123 & 16.123 $\pm$ 0.006 & 70.568 $\pm$ 0.009 & 19.209 $\pm$ 0.009 \\
-2.76$^{\circ}$ & 13.311 & 13.311 $\pm$ 0.035 & 58.620 $\pm$ 0.092 & 15.551 $\pm$ 0.034 \\
-3.25$^{\circ}$ & 11.038 & 11.038 $\pm$ 0.007 & 43.998 $\pm$ 0.012 & 12.670 $\pm$ 0.003 \\
-3.75$^{\circ}$ & 9.055 & 9.055 $\pm$ 0.003 & 30.954 $\pm$ 0.043 & 10.162 $\pm$ 0.014 \\
-4.25$^{\circ}$ & 7.494 & 7.494 $\pm$ 0.023 & 22.192 $\pm$ 0.034 & 8.146 $\pm$ 0.015 \\
-4.74$^{\circ}$ & 6.295 & 6.295 $\pm$ 0.010 & 16.486 $\pm$ 0.011 & 6.646 $\pm$ 0.003 \\
-5.23$^{\circ}$ & 5.488 & 5.488 $\pm$ 0.002 & 12.685 $\pm$ 0.036 & 5.632 $\pm$ 0.017 \\
-5.72$^{\circ}$ & 4.589 & 4.589 $\pm$ 0.003 & 8.919 $\pm$ 0.058 & 4.601 $\pm$ 0.030 \\
-6.23$^{\circ}$ & 3.934 & 3.934 $\pm$ 0.009 & 6.328 $\pm$ 0.007 & 3.809 $\pm$ 0.004 \\
\hline
\multicolumn{5}{c}{\textbf{DIA source}} \\
\hline
-1.40$^{\circ}$ & 21.931 & 21.931 $\pm$ 0.006 & 224.902 $\pm$ 1.242 & 25.553 $\pm$ 0.129 \\
-1.77$^{\circ}$ & 21.301 & 21.301 $\pm$ 0.009 & 261.791 $\pm$ 0.031 & 24.804 $\pm$ 0.014 \\
-2.26$^{\circ}$ & 17.846 & 17.846 $\pm$ 0.001 & 197.174 $\pm$ 0.049 & 20.509 $\pm$ 0.003 \\
-2.76$^{\circ}$ & 14.597 & 14.597 $\pm$ 0.040 & 150.158 $\pm$ 0.396 & 16.489 $\pm$ 0.040 \\
-3.25$^{\circ}$ & 12.024 & 12.024 $\pm$ 0.003 & 107.772 $\pm$ 0.135 & 13.364 $\pm$ 0.012 \\
-3.75$^{\circ}$ & 9.816 & 9.816 $\pm$ 0.030 & 73.464 $\pm$ 0.198 & 10.689 $\pm$ 0.028 \\
-4.25$^{\circ}$ & 8.054 & 8.054 $\pm$ 0.004 & 50.647 $\pm$ 0.039 & 8.542 $\pm$ 0.006 \\
-4.74$^{\circ}$ & 6.730 & 6.730 $\pm$ 0.001 & 36.212 $\pm$ 0.030 & 6.928 $\pm$ 0.006 \\
-5.23$^{\circ}$ & 5.786 & 5.786 $\pm$ 0.005 & 27.041 $\pm$ 0.007 & 5.819 $\pm$ 0.003 \\
-5.72$^{\circ}$ & 4.848 & 4.848 $\pm$ 0.019 & 18.826 $\pm$ 0.100 & 4.771 $\pm$ 0.027 \\
-6.23$^{\circ}$ & 4.153 & 4.153 $\pm$ 0.014 & 13.182 $\pm$ 0.010 & 3.958 $\pm$ 0.003 \\
\hline
\end{tabular}
}
\label{OMlistB}
\end{table*}

\subsection{O-C residual maps}
\label{sec:res}

\begin{figure*}
{\centering
\begin{tabular}{C{0.45\textwidth}C{0.52\textwidth}}
\textbf{Resolved sources} & \textbf{DIA sources} \\
\end{tabular}
\subfigure{\includegraphics[width=0.50\textwidth]{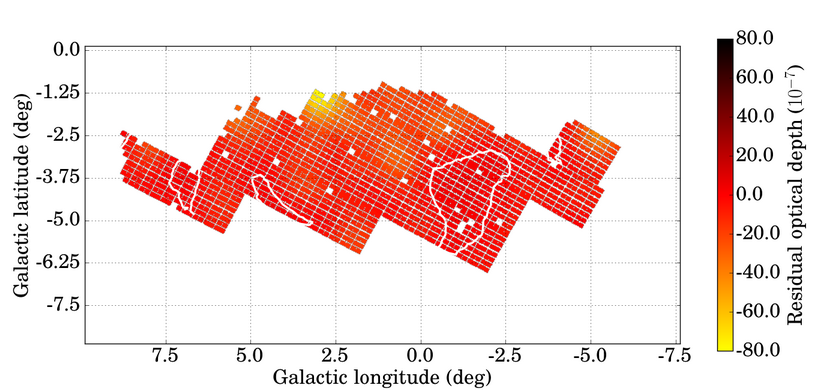}}\subfigure{\includegraphics[width=0.50\textwidth]{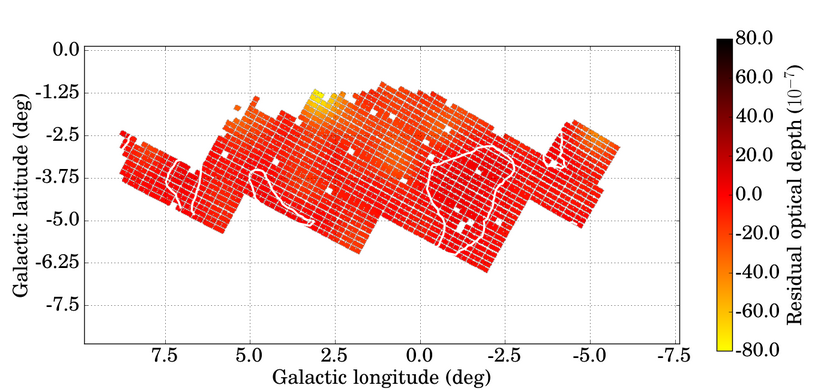}}
(a) Optical depth\\
\subfigure{\includegraphics[width=0.50\textwidth]{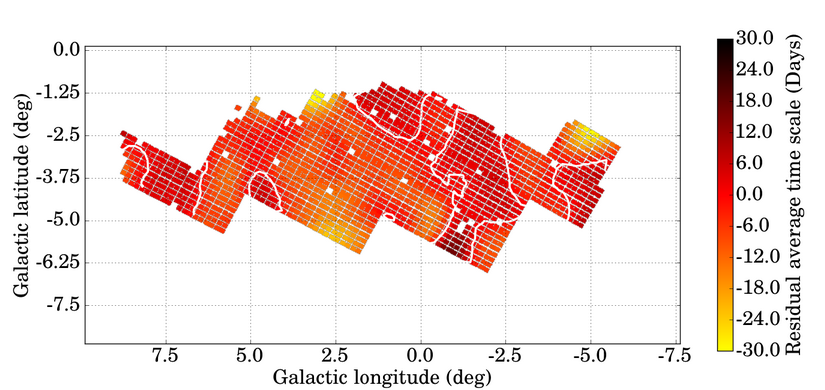}}\subfigure{\includegraphics[width=0.50\textwidth]{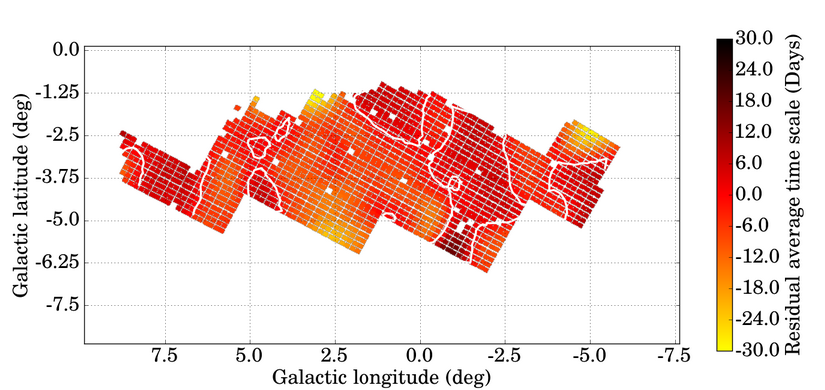}}
(b) Average time scale\\
\subfigure{\includegraphics[width=0.50\textwidth]{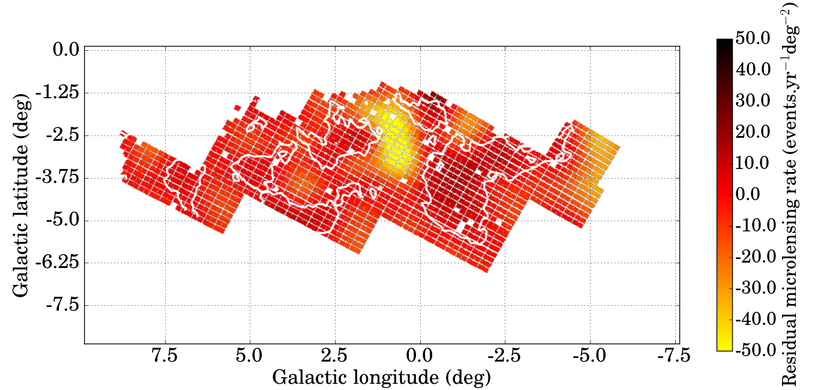}}\subfigure{\includegraphics[width=0.50\textwidth]{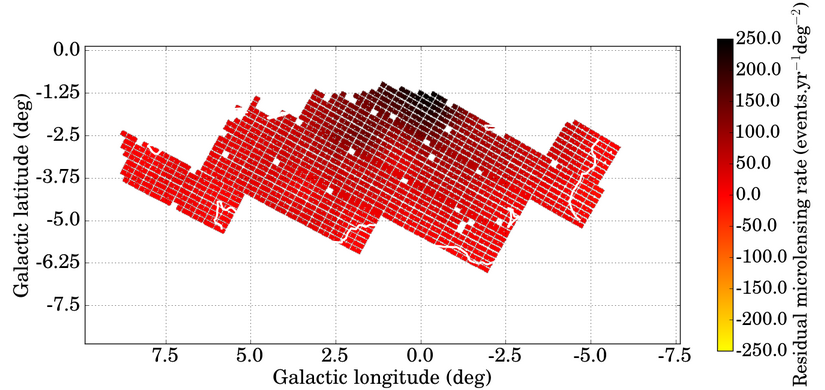}}
(c) Microlensing event rate per square degree\\
\subfigure{\includegraphics[width=0.50\textwidth]{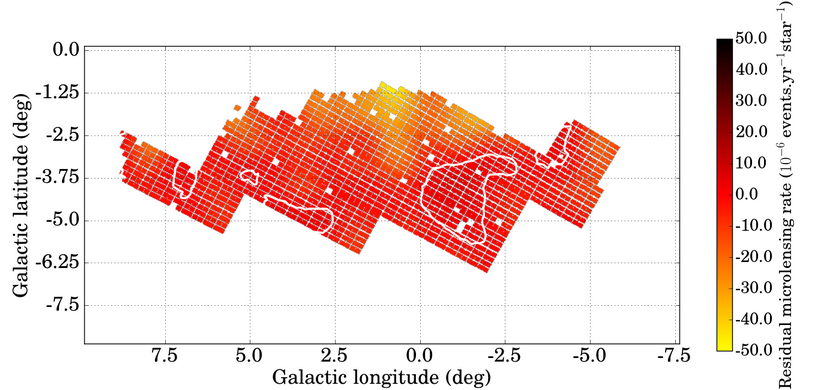}}\subfigure{\includegraphics[width=0.50\textwidth]{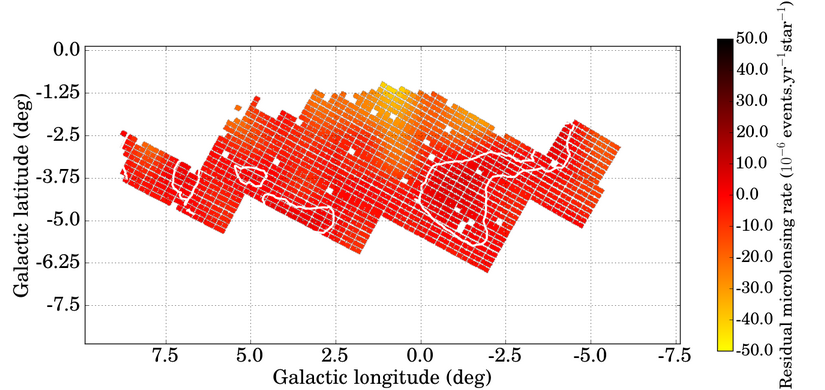}}
(d) Microlensing event rate per star\\
}
\caption[The optical depth, average time scale, microlensing event rate per square degree and microlensing event rate per star residual maps]{The optical depth (a), average time scale (b), microlensing event rate per square degree (c) and microlensing event rate per star (d) residual maps of resolved sources (left) and DIA sources (right) from the Besan\c{c}on Galactic model and the MOA-II survey data. Contour level shows zero residual value.}
\label{fig:res}
\end{figure*}

In Section~\ref{sec:besancon}, the model microlensing maps of the MOA-II field, filter and time scale cut are shown. In order to compare the result with the MOA-II observational data, residual maps are produced ($x_{Besancon}-x_{MOA}$). In Figure~\ref{fig:res}(a), the Besan\c{c}on model underestimates the optical depth compared with the MOA-II data at closer to the Galactic Centre ($b<3^{\circ}$), predicting only around $50\%$ of the MOA-II measured optical depth. However, moving away from the Galactic Centre, the Besan\c{c}on optical depth is generally in good agreement with the MOA-II measurement, suggesting that the Besan\c{c}on disk model provides a good description of the microlensing data.

The residual average time scale maps in Figure~\ref{fig:res}(b), show that the Besan\c{c}on model provide a reasonably good representation of the MOA-II timescales across pretty much the whole map. Most of the structures in the map are produced by individual very long time scale MOA-II events or many short time scale event in observed specific sub-fields. We therefore conclude that the {\em average}\/ microlensing kinematics within both the disk and bulge are consistent with microlensing data.

In Figures~\ref{fig:res}(c) and (d), which show the residual event rate per unit area and per source, respectively, indicate a similar deficit of the model with respect to the data within the inner bulge region.

These maps confirm the view that, whilst the model bulge kinematics provide a good average description of the event timescale, the mass is insufficient by a factor 2 to explain the observed number of events.

\subsection{Reduced $\boldsymbol{\chi^2}$ of $\boldsymbol{\tau}$ and $\boldsymbol{\Gamma_{\rm star}}$ maps}

\begin{figure*}
{\centering
\begin{tabular}{C{0.45\textwidth}C{0.52\textwidth}}
\textbf{Resolved sources} & \textbf{DIA sources} \\
\end{tabular}
\subfigure{\includegraphics[width=0.5\textwidth]{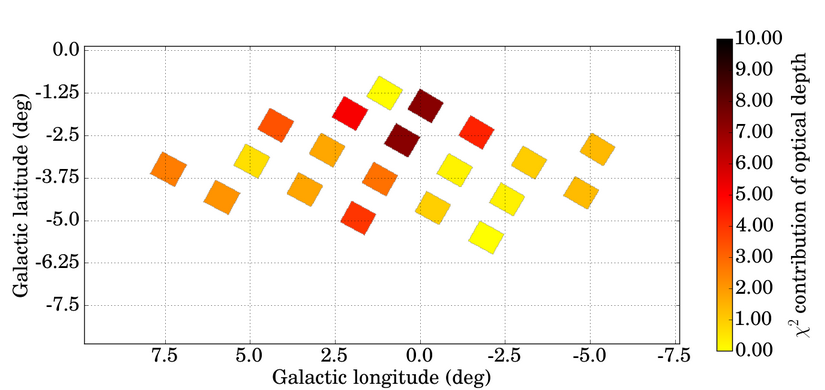}}\subfigure{\includegraphics[width=0.5\textwidth]{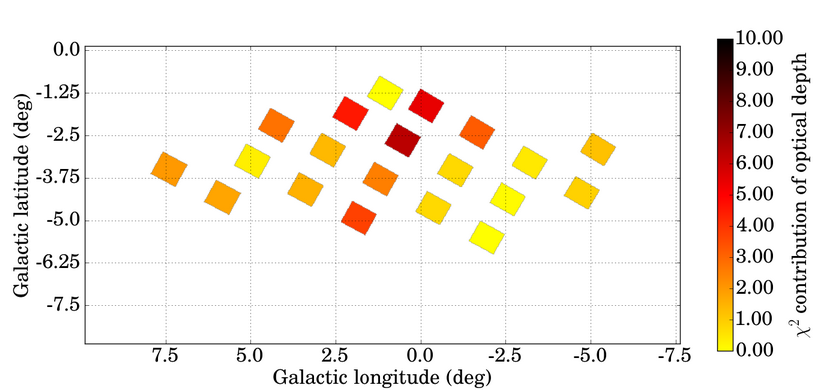}}
(a) Optical depth\\
\subfigure{\includegraphics[width=0.5\textwidth]{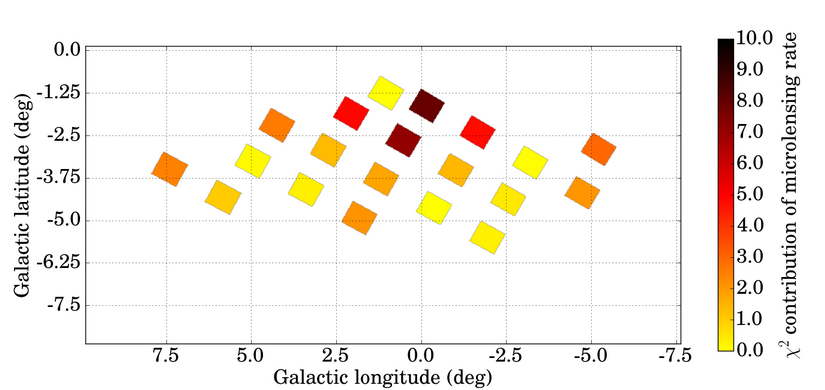}}\subfigure{\includegraphics[width=0.5\textwidth]{MOAMsuRC.png}}
(b) Microlensing event rate per star\\
}
\caption[The reduced $\chi^2$ contribution of optical depth and microlensing event rate per star]{The reduced $\chi^2$ contribution of optical depth (a) and microlensing event rate per star (b) of resolved sources (left) and DIA sources (right).}
\label{fig:field}
\end{figure*}

From Section~\ref{sec:res}, the simulated results under-predict the optical depth and microlensing event rate per star compared with the MOA-II observational data and show the structure at low Galactic latitude. The significance of this result can be assessed by a straightforward reduced $\chi^2$ statistic:
\begin{equation}
\chi^{2}_{r}=\frac{1}{N_{\textup{fld}}}\sum_{i=1}^{N_{\textup{fld}}} \left(\frac{x_{\textup{Bes},i}-x_{\textup{MOA},i}}{\sigma_{\textup{MOA},i}}
\right)^2,
\end{equation}
where
$(x_{\textup{Bes}},x_{\textup{MOA}})$ refer to the (model, observed) microlensing quantity. $\sigma$ is the observational uncertainty within the field, and $N_{\textup{fld}}$ is number of fields. The observational uncertainty within the field is calculated using formula from \cite{han1995}. The reduced $\chi^2$ contribution of each MOA-II field are shown in Table~\ref{fielddata} and Figure~\ref{fig:field}. The gb21 field is excluded due to the limit of the Besan\c{c}on extinction maps. The model optical depth is in agreement with MOA-II data within 3$\sigma_{\textup{MOA}}$ for most fields. The reduced $\chi^2$ of resolved source and DIA sources optical depth are 2.4 and 2.0, respectively.

The event rates show higher reduced chi-squared contribution than the optical depths. The Besan\c{c}on resolved sources and DIA source results have $\chi^2_r$ values of 2.6 and 2.2, respectively. The low Galactic latitude area ($b<3^{\circ}$) of both parameters provide the bulk of the disagreement (See Section~\ref{sec:opfit}).

In field gb1, there is a long time scale event (gb1-3-1, $t_{E}=157.6$ days) which contributes more than half of optical depth in that field. This event provides a hot spot in MOA-II optical depth and average time scale maps (Section~\ref{sec:res}). In order to check the reliability of reduced ${\chi^2}$ test, the reduced ${\chi^2}$ of field gb1 without gb1-3-1, gb1$_{\textup{Cut}}$, is calculated. The result in Table~\ref{fielddata} shows that the gb1$_{\textup{Cut}}$ field provide a better reduced $\chi^2_r$ than original gb1 field. 

Finally, we cut the events which have crossing time longer than 100 days which locate in 5 field; gb1, gb9, gb10, gb13 and gb 14. The new reduced chi-squared, $\chi^{2}_{r,\textup{Cut}}$, of optical depth (2.3 for resolved sources and 1.8 for DIA sources) and event rate per star (2.2 for resolved sources and 2.0 for DIA sources) is reduced. However, fields which locate around inner bulge except field gb6 still show high reduced chi-squared compare to high Galactic latitude field. Therefore, high reduced chi-squared region around inner bulge do not effect by long time scale event, but show the mismatch of optical depth and event rate per star between the Besan\c{c}on model and the MOA-II data at low Galactic latitude area. 
\begin{table}
\caption{Field-by-field contributions to the reduced $\chi^2$ ($\chi^2_r$) between the Besan\c{c}on model and the MOA-II data for the optical depth and microlensing event rate per star.}
\centering
{\footnotesize 
\begin{tabular}{lcccc}
\\
\hline\hline
Field & $\tau_\textup{Res}$ & $\tau_\textup{DIA}$ & $\Gamma_{\textup{star},\textup{Res}}$ & $\Gamma_{\textup{star},\textup{DIA}}$ \\ [3pt]
\hline
gb1 & 1.40 & 1.22 & 3.67 & 3.00 \\
gb2 & 1.34 & 0.91 & 2.47 & 2.09 \\
gb3 & 1.01 & 0.45 & 0.02 & 0.06 \\
gb4 & 4.31 & 3.24 & 5.91 & 4.78 \\
gb5 & 7.31 & 5.47 & 9.41 & 7.94 \\
gb6 & 0.04 & 0.01 & 0.15 & 0.02 \\
gb7 & 0.28 & 0.08 & 0.67 & 0.44 \\
gb8 & 0.26 & 0.77 & 0.84 & 1.44 \\
gb9 & 7.34 & 6.35 & 7.99 & 7.08 \\
gb10 & 5.14 & 4.56 & 5.76 & 4.91 \\
gb11 & $10^{-3}$ & 0.02 & 0.38 & 0.29 \\
gb12 & 0.97 & 0.75 & 0.03 & $10^{-3}$ \\
gb13 & 2.77 & 2.50 & 2.03 & 1.76 \\
gb14 & 1.72 & 1.39 & 1.74 & 1.34 \\
gb15 & 3.37 & 2.75 & 3.12 & 2.60 \\
gb16 & 3.98 & 3.73 & 2.36 & 2.13 \\
gb17 & 1.80 & 1.53 & 0.45 & 0.32 \\
gb18 & 0.64 & 0.32 & 0.27 & 0.12 \\
gb19 & 2.12 & 1.77 & 1.28 & 1.02 \\
gb20 & 2.56 & 2.02 & 2.91 & 2.49 \\
\hline
$\chi^{2}_{r}$ & 2.4 & 2.0 & 2.6 & 2.2 \\
\hline
gb1$_{\textup{Cut}}$ & 0.87 & 0.33 & 2.72 & 2.10 \\
gb9$_{\textup{Cut}}$ & 7.34 & 6.34 & 7.72 & 6.83 \\
gb10$_{\textup{Cut}}$ & 4.20 & 3.34 & 4.35 & 3.57 \\
gb13$_{\textup{Cut}}$ & 1.89 & 1.57 & 1.50 & 1.26 \\
gb14$_{\textup{Cut}}$ & 0.77 & 0.40 & 1.05 & 0.72 \\
\hline
$\chi^{2}_{r,\textup{Cut}}$ & 2.3 & 1.8 & 2.4 & 2.0 \\
\hline
\end{tabular}
}
\label{fielddata}
\end{table}

\section{Microlensing model parametrisation}

\begin{figure*}
{\centering
\begin{tabular}{C{0.45\textwidth}C{0.52\textwidth}}
\textbf{Resolved sources} & \textbf{DIA sources} \\
\end{tabular}
\subfigure{\includegraphics[width=0.50\textwidth]{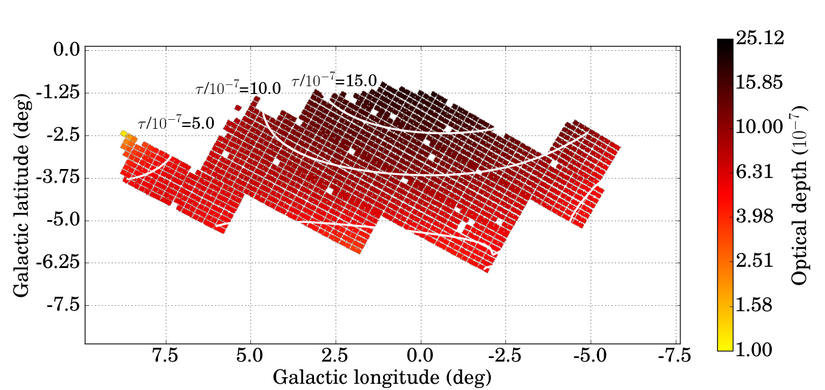}}\subfigure{\includegraphics[width=0.50\textwidth]{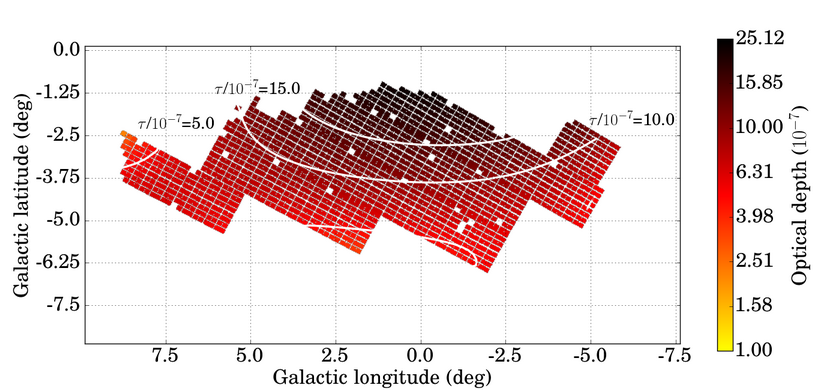}}
(a) Optical depth\\
\subfigure{\includegraphics[width=0.50\textwidth]{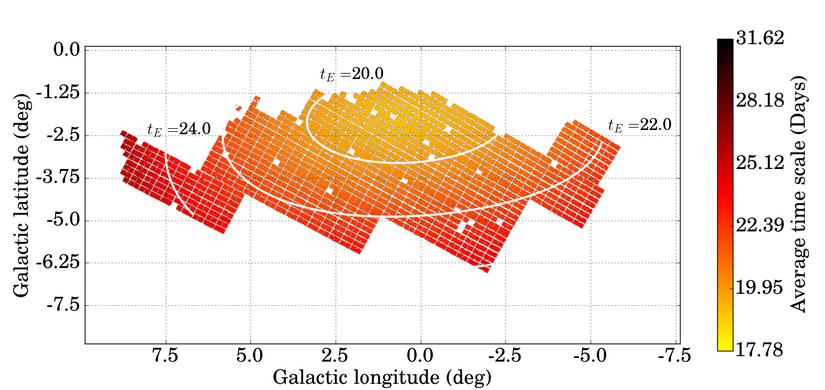}}\subfigure{\includegraphics[width=0.50\textwidth]{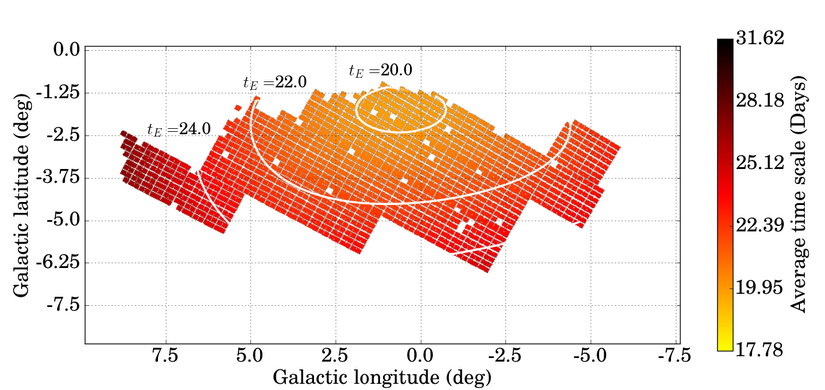}}
(b) Average time scale\\
\subfigure{\includegraphics[width=0.50\textwidth]{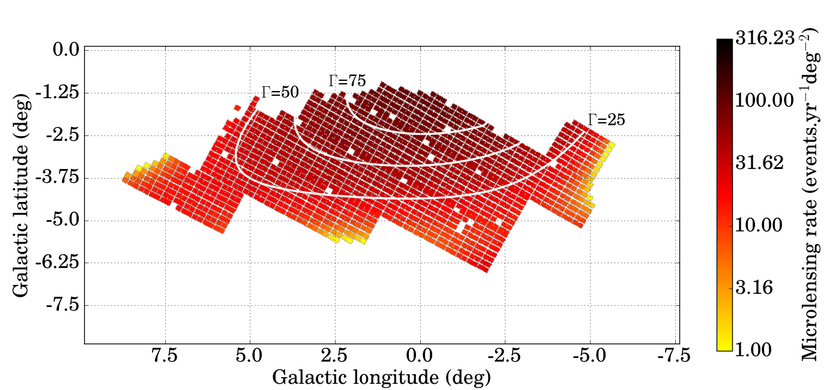}}\subfigure{\includegraphics[width=0.50\textwidth]{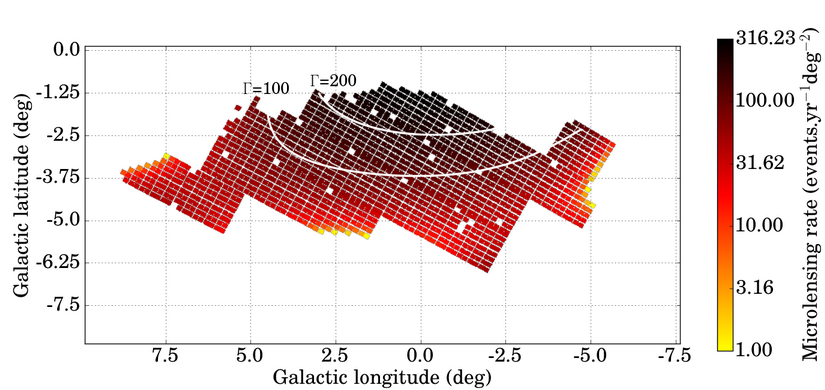}}
(c) Microlensing event rate per square degree\\
\subfigure{\includegraphics[width=0.50\textwidth]{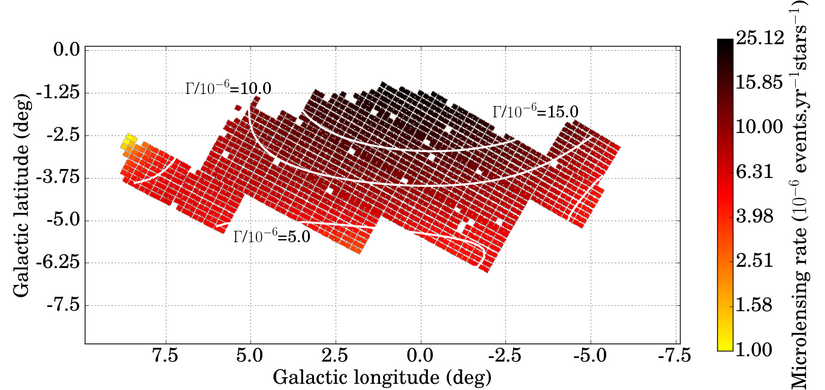}}\subfigure{\includegraphics[width=0.50\textwidth]{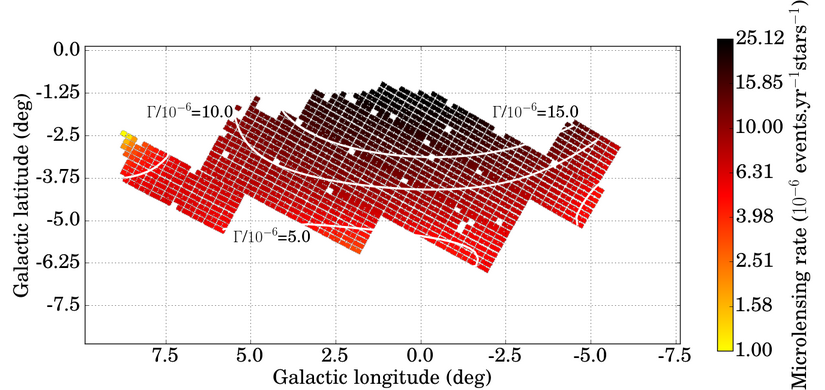}}
(d) Microlensing event rate per star\\
}
\caption[Optical depth, average time scale, microlensing event rate per square degree and microlensing event rate per star best fit model maps]{Best-fit parametrised representations of the Besan\c{c}on model maps shown in Figure~\ref{fig:map}. Best fit parametrisations are shown for the optical depth (a), average time scale (b), microlensing event rate per square degree (c) and microlensing event rate per star (d) for resolved sources (left) and DIA sources (right). The parameters of the fits are given in Table~\ref{Model}.}
\label{fig:fit}
\end{figure*}

The MOA-II team parameterise the observed spatial microlensing distribution using a polynomial function. We can do likewise for our simulated maps. We model the structure of the optical depth, average time scale and event rate maps shown in Figure~\ref{fig:map} using a 10-parameter cubic polynomial fit in $l$ and $b$. The model function can be written as,
\begin{equation}
x=a_{0}+a_{1}l+a_{2}b+a_{3}l^{2}+a_{4}lb+a_{5}b^{2}+a_{6}l^{3}+a_{7}l^{2}b+a_{8}lb^{2}+a_{9}b^{3} \ ,
\end{equation}
where $x$ is the microlensing observable (rate, time-scale or optical depth). The best-fit models are shown in Figure~\ref{fig:fit} and the model parameters are provided in Table~\ref{Model}. The best fit models agree to within 20\% of the exact model value for $|b|<5^{\circ}$. 

\begin{table*}
\caption{The best fit model parameter of the Besan\c{c}on microlensing parameters.}
\centering
{\footnotesize 
\begin{tabular}{lcccccccc}
\\
\hline\hline
& $\tau_\textup{Res}$ & $\tau_\textup{DIA}$ & $\left \langle t \right \rangle _\textup{Res}$ & $\left \langle t \right \rangle _\textup{DIA}$ & $\Gamma_{\textup{deg}^{2},\textup{Res}}$ & $\Gamma_{\textup{deg}^{2},\textup{DIA}}$ & $\Gamma_{\textup{star},\textup{Res}}$ & $\Gamma_{\textup{star},\textup{DIA}}$ \\ [3pt]
& $\times 10^{-8}$ & $\times 10^{-8}$ & days & days & yr$^{-1}$deg$^{-2}$ & yr$^{-1}$deg$^{-2}$ & $\times 10^{-7}$yr$^{-1}$star$^{-1}$ & $\times 10^{-7}$yr$^{-1}$star$^{-1}$ \\ [3pt]
\hline
$a_{0}$ & 256 & 321 & 21.7 & 21.5 & 101 & 477 & 305 & 378 \\
$a_{1}$ & -14.8 & -16.9 & -0.151 & -0.260 & -11.0 & -38.9 & -17.2 & -19.0 \\
$a_{2}$ & 35.8 & 63.5 & 2.63 & 1.91 & -14.5 & 111 & 38.4 & 75.1 \\
$a_{3}$ & -4.21 & -4.52 & 0.121 & 0.126 & -3.99 & -10.3 & -5.63 & -5.89 \\
$a_{4}$ & -7.85 & -8.70 & 0.012 & -0.043 & -7.49 & -22.9 & -9.42 & -10.1 \\
$a_{5}$ & -3.84 & 0.464 & 0.817 & 0.631 & -12.9 & -1.02 & -6.93 & -0.191 \\
$a_{6}$ & 0.087 & 0.104 & 0.002 & 0.003 & 0.053 & 0.191 & 0.112 & 0.127 \\
$a_{7}$ & -0.720 & -0.771 & 0.016 & 0.016 & -0.735 & -1.88 & -0.965 & -1.00 \\
$a_{8}$ & -1.02 & -1.12 & 0.004 & -0.004 & -1.08 & -3.13 & -1.25 & -1.32 \\
$a_{9}$ & -0.647 & -0.408 & 0.056 & 0.042 & -1.31 & -1.10 & -1.00 & -0.570 \\
\hline
\end{tabular}
}
\label{Model}
\end{table*}

\section{Conclusion}

A new version of the Besan\c{c}on Galactic model is used to simulated microlensing optical depth, average timescales and microlensing event rate maps towards the Galactic bulge. The new model incorporates a refined two-component bulge \citep{rob2012}. We perform a detailed comparison of the model with the recent optical depth study by MOA-II \citep{sum2013} based on 474 events. The MOA-II observational filter, time scale cut and Gaussian kernel are applied to the maps. This is the first detailed field-by-field comparison between a theoretical microlensing model and a large-scale microlensing dataset. 

In its original form the model overestimates the average time scale compared to the survey because the model lacks low-mass stars. Allowing for an extension of the model stellar mass function into the low mass star and brown dwarf regime, we find that the model correctly produces the observed average event timescale provided the mass function is essentially cut off at the hydrogen burning limit. The shape of the observed timescale distribution shows weak evidences of an excess of short ($0.3<t_{E}<2$ days) and very long ($30<t_{E}<200$ days) duration events and a deposit of moderate duration events ($2<t_{E}<30$ days). However, the model provides satisfactory match with MOA-II distribution (reduced $\chi^2 \simeq 2.2$).

Encouragingly, the inferred efficiency corrected MOA-II event rate is found to lie between the predicted number of events from the Besan\c{c}on model for pure resolved sources and DIA sources. The number of Besan\c{c}on microlensing events with resolved sources and DIA sources are 0.83 and 2.17 times number of MOA-II detected events. Given that the model analysis does not include a correction for blending in the number of available sources, and some expected differences due to differences in the assumed bandpass, this is a reasonable level of agreement.

For the optical depth the residual maps between the model predictions and MOA-II observations show that there is generally good agreement over most of the MOA-II survey area and that the disagreement is confined to the regions closest to the Galactic Centre ($b<3^{\circ}$). The Besan\c{c}on model predicts only 50\% of the observed optical depth in this region. Maps of the event rate per star maps also show a similar disagreement. The fact that there is reasonable agreement in the maps of average duration but disagreement with the rate and optical depth argues for a mass deficit in the current bulge model.

The bulge mass employed in the current Besan\c{c}on model is somewhat lower than inferred in some recent studies such as \cite{por2015}, who argue that a bar with a mass in the range $(1.25-1.6)\times10^{10}~M_{\odot}$ is compatible with recent radial velocity and proper motion studies. Such a massive bar is likely required to solve the optical depth discrepancy reported here. It remains to be seen whether such a model can be straightforwardly accommodated within a full population synthesis code. However, \cite{rob2012} argue that the dust map model is likely to be under-estimated in the innermost regions due to incompleteness of 2MASS star counts below $K \simeq 12$. They also identify a missing population within the inner $\sim 1^{\circ}$ in their model based on star count residuals. The additional population, along with increased extinction in this region should permit an increased optical depth without violating star count limits. Further work, perhaps using VISTA data to extend the range of the dust map in the innermost regions of the bulge, should enable us to refine the Besan\c{c}on model. 

We are now in an era of large-scale microlensing datasets which will play a pivotal role in our understanding of the inner structure of the Galaxy. There is no shortage of pieces to be included within the Galactic jigsaw, but it remains a testing puzzle to assemble them into a fully synthesized view of our Galaxy.

\section*{Acknowledgments}

The authors acknowledge the anonymous referees for their valuable suggestions that have helped to improve this paper. The authors are also grateful to Takahiro Sumi for providing the MOA-II survey data. SA gratefully acknowledges the support from the Thai Government Scholarship and the University of Manchester President's Doctoral Scholar Award.

\bibliographystyle{mnras} 
\bibliography{MNRASrefs}

\appendix

\bsp

\label{lastpage}

\end{document}